\documentstyle[aps,prb,epsf,preprint]{revtex}

\begin{document}
\title{Approximation of excitonic absorption in disordered systems
using a compositional component weighted
CPA}
\draft
\author{N.F. Schwabe and R.J. Elliott}
\address{Department of Physics, University of Oxford \\
Theoretical Physics, 1 Keble Road, Oxford OX1 3NP, UK}
\date{\today}
\maketitle

\begin{abstract}
Employing a recently developed technique of component weighted two
particle Green's functions in the CPA of a binary substitutional alloy
$A_cB_{1-c}$ we extend the existing theory of excitons in such media
using a contact potential model for the interaction between
electrons and holes to an approximation which
interpolates correctly between the limits of weak and strong
disorder. With our approach we are also able to treat the case where
the contact interaction between carriers varies between sites of
different types, thus introducing further disorder into the system.
 Based on this approach we study numerically how the formation of
exciton bound states changes as the strengths of the contact
potentials associated with either of the two site types are varied
through a large range of parameter values.
\end{abstract}
\vspace{3ex}

\section{Introduction}

Excitonic optical absorption in strongly disordered semiconductor
 alloys has been
of great interest to both semiconductor physics and technological
applications of semiconductors and semiconductor structures in the
past. Although the theory of an
optically excited electron-hole pair scattered or bound under the
 influence of the
mutual Coulomb interaction into an exciton is quite
well understood in weakly disordered systems and even at finite
 carrier
densities in the respective conduction and valence bands, theories for
strongly disordered systems such as alloys of insulators and some
semiconductors are still very incomplete.

One of the few theories of disorder
which interpolates correctly
between all regimes of disorder strengths and concentrations is the
 coherent potential
approximation CPA first developed by Soven \cite{bi4} and Taylor
 \cite{bi6}
and subsequently extended to a proper two particle theory by
Velicky \cite{bi10}. The CPA predicts correctly the occurrence of
split-off impurity bound states and whole impurity bands,
 once the relative disorder strength becomes of the order of the
width of the
unperturbed single particle density of states considered.
Despite the numerous advantages that the
CPA provides, it has proved difficult to
incorporate a treatment of a carrier-carrier interaction into its
framework
and to our knowledge the only attempt to find a joint treatment of
 both effects has
been made by Kanehisa and Elliott \cite{bi2} who introduced a random
phase like decoupling of the disorder average in the corresponding
perturbation expansion in combination with a contact potential model
 for
the Coulomb interaction. Although this description produces a number
 of correct
features in a limit of low disorder, it could not be extrapolated
successfully to the case where the disorder becomes stronger, i.e.
 when the joint
density of states splits into two components and the excitons move to
a more localized Frenkel limit, where a separate approach is
needed. No theory has been available for the region of
intermediate disorder strengths where the bands are about to split.

 An improved treatment of
this problem has become possible through the development of a
properly weighted two particle CPA in a previous paper \cite{bi12}
which allows to distinguish between different site types involved in
the propagation of two particles during an absorption process. We will
show that differentiating between alloy components in a perturbation
expansion with an appropriate decoupling procedure of the disorder
average can yield the correct interpolation to all ranges of
disorder for a contact potential model.
Furthermore, this allows us to distinguish different
contact interaction strengths between the carriers if these meet on
different types of sites, whereby these strengths are modeled
to represent the
electron-hole interaction within an atomic radius.

\section{Absorption in the non-interacting disordered alloy}

\subsection{CPA Model of the Disordered System}

We summarize here the results for a component  weighted two particle
CPA.
Our model system is a binary substitutional alloy of components
$A$ and $B$ with respective concentrations $c$ and $1-c$, on a simple
 three dimensional monoatomic lattice. Both components are assumed to
 exhibit a direct band
gap in their pure phase.
Our two-band model Hamiltonian without carrier-carrier interactions
 can thus be written in a site
representation as
\begin{equation}
H= \sum_{l,n} \{ W^c_l c_{n+l}^+ c_n + W^v_l d_{n+l}^+ d_n \} + \sum_n
\{ U^c_n c_n^+ c_n +  U^v_n d_n^+ d_n \} \label{m1}
\end{equation}
where the $c_n$ and $d_n$ are the annihilation operators for an
 electron
and a hole on the site $n$, respectively. $W^{\mu}$ is the periodic
 part of the
Hamiltonian transferring particles between different sites in the band
$\mu$ and
$U_n^{\mu}$ is the matrix element in the respective bands which
assumes
the values $U^{\mu}_n \in \{
\varepsilon_A^{\mu}, \varepsilon_B^{\mu} \}$ depending on whether $n$
is an $A-$ or a $B$-site. The disorder in the system is generated by
the difference of the on-site energies
$V^{\mu}=\varepsilon_A^{\mu}-\varepsilon_B^{\mu}$ in the respective
 bands.

The CPA for a disordered medium is introduced by the usual method
\cite{bi4,bi6,bi8} of placing the impurities in a  self-consistent
medium such that the scattering off a single impurity vanishes on the
average.
Independent of this approximation, the single particle propagator of
 the disordered
medium $G$ relates to the one of a pure medium
$g$ through the Dyson equation
\begin{equation}
G^{\mu}=g^{\mu}+g^{\mu}U^{\mu}G^{\mu} \label{m2}.
\end{equation}
The the self consistent CPA condition requires that the average
  propagator
$ \bar G^{\mu} $ of the effective medium fulfill the relation
\begin{equation}
\bar G^{\mu}=g^{\mu}+g^{\mu} \Sigma^{\mu} \bar G^{\mu}. \label{m3}
\end{equation}
where $\Sigma^{\mu}$ is the CPA single particle self energy of
the effective medium.
$\Sigma^{\mu}$ itself is determined through the single site condition
 that the
average atomic T-matrix of the self consistent medium $T^{\mu}_n$
 defined as
\begin{equation}
T^{\mu}_n =  (U_n^{\mu} - \Sigma^{\mu}) + (U_n^{\mu} - \Sigma^{\mu})
F^{\mu} T^{\mu}_n\label{m4},
\end{equation}
be zero. Here we have introduced the site diagonal single particle
 function
$F= \bar G(n,n)$. $\Sigma^{\mu}$ and $F^{\mu}$ therefore satisfy the
self-consistent relation
\begin{equation}
0 \equiv \langle T^{\mu}_n \rangle = \frac { c[\varepsilon_A^{\mu} -
\Sigma^{\mu}]} {1- [\varepsilon_A^{\mu} -
 \Sigma^{\mu}]F^{\mu}} + \frac { (1-c)[\varepsilon_B^{\mu} -
\Sigma^{\mu}]} {1- [\varepsilon_B^{\mu} - \Sigma^{\mu}]F^{\mu}}
\label{m5}.
\end{equation}

Weights can now be attributed to the single particle functions before
averaging through applying operators $\Xi^{i/h}_{\mu}$ to them which
effectively exclude either impurity or host sites from the choices of
start or end sites of the propagation of the particles. We make the
following definitions
\begin{equation}
\begin{array}{ccccccc}
 \Xi^A_{\mu}= [U_n^{\mu}-\varepsilon_B^{\mu}]/V^{\mu} ; &
 \hspace{5mm} &G^A =
\Xi^A_{\mu} G ;& \hspace{5mm} &G^{AA} = \Xi^A_{\mu} G_{\mu}
\Xi^A_{\mu} ; & \hspace{5mm} &G^{AB} = \Xi^A_{\mu} G_{\mu}
 \Xi^B_{\mu} , \\
\Xi^B_{\mu}= [\varepsilon_A^{\mu}-U_n^{\mu}]/V^{\mu} ; &
\hspace{5mm} &G^B =
\Xi^B_{\mu} G ;  & \hspace{5mm} &G^{BB} = \Xi^B_{\mu} G_{\mu}
\Xi^B_{\mu} ; &
\hspace{5mm} & G^{BA} = \Xi^B_{\mu} G_{\mu}  \Xi^A_{\mu} , \\
\end{array} \label{m6}
\end{equation}
Averaged versions of such weighted single particle functions have
 been
 introduced by Aiyer et al.\cite{bib} It will be shown that
 they are
particularly useful in their unaveraged form when employed
 in the definitions of two particle functions in order to
calculate weighted components of the polarizability which can
 thereupon be
used in the treatment of the excitonic absorption.

\subsection{The Average Polarizability in Linear Response Theory}

In the following we consider a linear response expression for
the polarizability of the Kubo type. The particular kind of two
 particle function needed in this case is
determined by the form of the dipole operators which account for
the interaction of the electronic excitation with the radiation field
and which in the case of allowed interband transitions \cite{bi7}
 effectively couple the two single particle resolvents in the
respective bands:
\begin{equation}
{\bf \hat p^{\dag}} = \sum_n \mid nc \rangle p \langle nv \mid .
 \label{mb1}
\end{equation}
The total optical polarizability in a Matsubara representation which
we adopt for convenience can be written as
\begin{equation}
\langle \Pi_0( i \omega ) \rangle = - \beta^{-1} \sum_{iz} \langle
Tr_v \{ {\bf \hat p} G_c(iz) {\bf \hat p^{\dag}} G_v(i\omega - iz) \}
 \rangle \label{mb2},
\end{equation}
where the angular brackets denote the configurational average, $\beta$
is the usual inverse temperature and $Tr_v$ denotes the trace over the
band $v$ as in Ref. \onlinecite{bi11}
\begin{equation}
Tr_v( \cdots )= \sum_n \langle n v \mid \cdots \mid n v
\rangle. \label{mb3}
\end{equation}
Here it is assumed that the dipole matrix elements are taken for the
 transition from a
valence band of $p$-wave symmetry to a conduction band of $s$-wave
 symmetry
in which case they can be assumed to be essentially constant and we
normalize them
to unity. This implies that for the calculation of a properly averaged
polarizability which is void of any further interactions
we need to consider the following two particle function
\begin{equation}
K(z_1,z_2)= \sum_m K(z_1,z_2;n,m)= \sum_m \langle nc \mid \langle
 G_c(z_1) \mid mc \rangle \langle
mv \mid G_v(z_2) \rangle \mid nv \rangle \label{mb4}.
\end{equation}
Apart from the cumulative function $K$ stated above it is now also
possible to consider its weighted components by using the
appropriately weighted versions of the single particle functions in
their definitions.

$K$ and its weighted forms are obtained by means of the
two particle extension of the CPA by  Velick{\'y}\cite{bi10}
 and our method of obtaining the appropriate weights\cite{bi12}.
The concept is to use the T-matrix
representation of the unaveraged unweighted or weighted single
particle functions from (\ref{m6}) to obtain a CPA vertex correction
$\Lambda(z_1,z_2)$ and average weights for the
correlated motion of two particles in terms of averages over products
of atomic T-matrices. With the vertex correction,
the equation for the correlated two particle motion can be decoupled
 as
\begin{equation}
K(z_1,z_2)= \frac { R(z_1,z_2) }{1 - \Lambda(z_1,z_2) R(z_1,z_2)} ,
 \label{mb5}
\end{equation}
where $R(z_1,z_2)$ is the average-decoupled two particle function
\begin{equation}
R(z_1,z_2) = \sum_m \bar G_c(z_1;n,m) \bar G_v(z_2;m,n) \label{mb6}
\end{equation}
and the vertex correction is found to be
\begin{equation}
\Lambda(z_1,z_2) = \frac {V_v \Sigma_c(z_1) - V_c \Sigma_v(z_2)}{V_c
F_c(z_1) - V_v F_v(z_2)}. \label{mb7}
\end{equation}
If one assumes that the conduction and valence band dispersions are
similar in shape
\begin{equation}
\frac {\varepsilon_c(k)}{w_c} = \mp \frac {\varepsilon_v(k)}{w_v}
 \label{mb6_2},
\end{equation}
one finds a more explicit representation for $R(z_1,z_2)$ \cite{bi11}
\begin{equation}
R(z_1,z_2) = \frac {w_c F_c(z_1) \pm w_v F_v(z_2)} {w_c [z_2 -
\Sigma_v(z_2)] \pm w_v [z_1 -\Sigma_c(z_1)]} \label{mb6_3}.
\end{equation}
In Ref. \onlinecite{bi12} we found that for the class of two particle
 functions
as represented by $K(z_1,z_2)$
two types of single weighted functions and three different double
weighted ones exist.
 Similar to the single particle
theory it was found that these weighted two particle functions can be
expressed in terms of the unweighted one if energy dependent weighting
factors added to and multiplied onto it.

 Introducing
\begin{eqnarray}
\xi(z_1,z_2) & = & \frac {(1-c) \Lambda(z_1,z_2)}
{(V_c-\Sigma_c(z_1))(V_v-\Sigma_v(z_2))} , \label{mb7_1}\\
\eta(z_1,z_2) & = & \frac {c \Lambda(z_1,z_2)}
{\Sigma_c(z_1)\Sigma_v(z_2)} , \label{mb7_2}\\
\gamma(z_1,z_2) & = & \frac
{\Lambda(z_1,z_2) F_c(z_1) F_v(z_2)} {V_c V_v},
 \label{mb7_3}
\end{eqnarray}
we find that
\begin{equation}
\begin{array}{ccccc}
K^A = \xi K ; & \hspace{5mm} & K^B = \eta K  , & & \\
K^{AA} = \xi^2 K + \gamma ; & \hspace{5mm} & K^{BB} = \eta^2 K +
\gamma ; & \hspace{5mm} &
 K^{AB} =  K^{BA} = \xi \eta K - \gamma . \\
\end{array} \label{mb8}
\end{equation}
These functions can now readily be used to obtain corresponding
components of the polarizability which are calculated the same way as
in (\ref{mb2})
\begin{equation}
\langle \Pi_0^{XY}(i\omega) \rangle = - \beta^{-1} \sum_{iz}
 K^{XY}(iz,i\omega
- iz) \label{mb10},
\end{equation}
where $X$ and $Y$ denote possible weights $(X,Y \in
 \{A,B,\emptyset \})$.
Functions of this type have been discussed analytically and numerically
in Ref. \onlinecite{bi12} for various regimes of disorder.

\section{Introduction of electron-hole interaction}

\subsection{ Difficulties with the Treatment of the Coulomb
Interaction in Disordered Systems}

The inclusion of a carrier-carrier interaction into a model for the
optical polarizability is essential for the treatment of exciton
 effects. However, an analytic treatment of this problem in strongly
disordered systems such as alloys of insulators and of semiconductors
using the true or screened
Coulomb interaction seems to be
almost impossible.
This is due to the fact that the corresponding Bethe-Salpeter equation
\begin{equation}
\langle {\cal K}(i \omega;i,j;k,l) \rangle  =
\langle {\cal K}_0(i \omega;i,j;k,l)
\rangle + \sum_{a,b,c,d}
\langle {\cal K}_0(i \omega;i,j;a,b) u(a,b;c,d)
{\cal K}(i \omega;c,d;k,l) \rangle \label{ea1}.
\end{equation}
is extremely hard to decouple. Here $u(a,b;c,d)=v(a-b) \delta_{ac}
\delta_{bd} - w(a-c) \delta_{ab} \delta_{cd}$ is the Coulomb
interaction in the site representation with its direct and exchange
part and
${\cal K}_0 (i \omega;i,j;k,l)$ is the non-interacting
configuration dependent two particle function
\begin{equation}
{\cal K}_0(i \omega;i,j;a,b) = - \beta^{-1} \sum_{i \omega}
\langle i \mid
G_c(iz) \mid k \rangle \langle l \mid G_v(iz - i \omega) \mid j
\rangle.
\end{equation}
The exactly averaged polarizability including exciton effects is
calculated from the interacting version of
this function (\ref{ea1}) as
\begin{equation}
\langle \Pi( i \omega) \rangle = \sum_{i,k} \langle {\cal K}(i
\omega;i,i;k,k) \rangle . \label{ea2}
\end{equation}
The main difficulty in solving equation (\ref{ea1}) arises from the
 fact that
the disorder average over the second term creates higher order
correlations in the particle motion than those which are accounted
 for by the
inclusion of the vertex corrections. These additional average
 induced correlations
lead to an effective Coulomb interaction which is renormalized in a
very complicated fashion.

For these reasons two drastic simplifications
have been made in earlier treatments of the problem. First the Coulomb
interaction is replaced by a short range contact potential which binds
at most one state in a joint band limit when the disorder is
relatively weak. This state is meant to represent the 1-$s$ exciton
 from
the Coulomb series which usually
dominates the optical absorption spectrum below the continuum
absorption edge. Secondly the
disorder average in (\ref{ea1}) is decoupled into a product of
averages. The average polarizability including this carrier-carrier
interaction is then obtained as
\begin{equation}
\bar \Pi( i \omega)  = \frac {\bar \Pi_0( i \omega) } {1 + u \bar
\Pi_0( i \omega)} \label{ea2_2},
\end{equation}
where $u$ is the uniform strength of the contact potential and the
 average was replaced through a horizontal bar over corresponding
quantities.
However, this type of decoupling of the average
becomes reasonable only in the limit of weak disorder, when the
electron-hole pair experiences many impurities before its
 constituents are Coulomb
scattered (Wannier excitons). Effectively, therefore no higher than
two particle correlations can be accounted for within this
approximation and it will be shown that for stronger disorder
even these correlations are significantly misweighted and
 overcounted. It hence must
clearly fail in a regime of strong disorder where excitons may be
 primarily
associated with two separate impurity bands.

For the strong
disorder limit it is shown explicitly in Ref. \onlinecite{bi2}
 how the solution for the two particle function
is modified towards a regime of
 localized excitations (Frenkel excitons),
which is equivalent to letting the overlap of the atomic
wavefunctions go to zero, while at the same time the
disorder strengths are kept constant.
If (\ref{ea1}) is considered before averaging in this limit,
it goes over to
\begin{equation}
[1 + \pi_0(i \omega;i)v(0)]{\cal K}(i \omega;ii,kk )=
\delta_{ik} \pi_0(i \omega;i) + \pi_0(i
\omega;i) \sum_j w(i-j){\cal K}(i \omega;jj,kk) \label{ea3},
\end{equation}
where $\pi_0(i \omega;i)$ is the atomic polarizability on the site $i$
and $w(i-j)$ is the exchange part of the Coulomb interaction which
 is now entirely responsible for electron transfer from one atom to
another. It is this term which makes
an exact solution of equation (\ref{ea3}) still difficult. If now a
the exchange $w$ is assumed
to go to zero as well we obtain the limit of isolated atoms.

The solution of (\ref{ea3}) then becomes
\begin{equation}
{\cal K}(i \omega;ii,kk) = \frac {\delta_{ik} \pi_0(i \omega,i)} {1 +
v(0)\pi_0(i\omega,i)} \label{ea4},
\end{equation}
which can be averaged exactly and summed to yield
\begin{equation}
\bar \Pi(i \omega) = \left[ \frac {c \pi_0(i\omega,A)} {1 +
v(0)\pi_0(i\omega,A)} + \frac {(1-c) \pi_0(i\omega,B)} {1 +
v(0)\pi_0(i\omega,B)} \right], \label{ea5}
\end{equation}
where $A$ and $B$ within $\pi_0$ denote here that the atomic
polarizabilities are considered on either an
$A$ or a $B$ atom.
Since the present limit corresponds to letting the bandwith of the
 components of the joint density of states go to zero, given some
arbitrary strength of disorder, the same limiting behavior of the
 polarizability can be attained \cite{bi12} by letting the disorder
 strengths
become much larger than the band width which yields
two energetically separated contributions to the joint DOS and the
 formal properties of (\ref{ea5})
are recovered.

The appearance of (\ref{ea5}) suggests that one
should  also allow for the possibility of using two different
contact
interaction strengths on different types of sites thus allowing for
 different
types of atomic screening in different alloy components.

It is clear that equation (\ref{ea5}) is not obtained by the
extrapolation to a strong disorder limit from equation (\ref{ea2_2})
 and so far
there had been no indication as to what type of approximation one
 should use in
the description of an intermediate disorder regime, where the
 strengths of the disorder
$V_{\mu}$
are of the order of the corresponding half-widths of the single
 particle bands $w_{\mu}$, which at the same time
 extrapolates correctly to the
 asymptotic behavior predicted by
equations (\ref{ea2_2}) and (\ref{ea5}) for weak and strong disorder,
 respectively.

\subsection{Component Weighted Approximation of the Scattering
 Expansion}

Based on our results in Ref. \onlinecite{bi12} we employ in the
 following a
 component weighted scattering expansion to obtain the excitonic
polarizability, while at the same time we allow the contact potential
strengths to assume different values on different types of sites.
 We use a
similar decoupling of the disorder averages to obtain a new
approximation for the scattering in an intermediate disorder regime.
Subsequently, we show that this approximation renders exactly the
 interpolation
behavior that we had hoped to achieve.

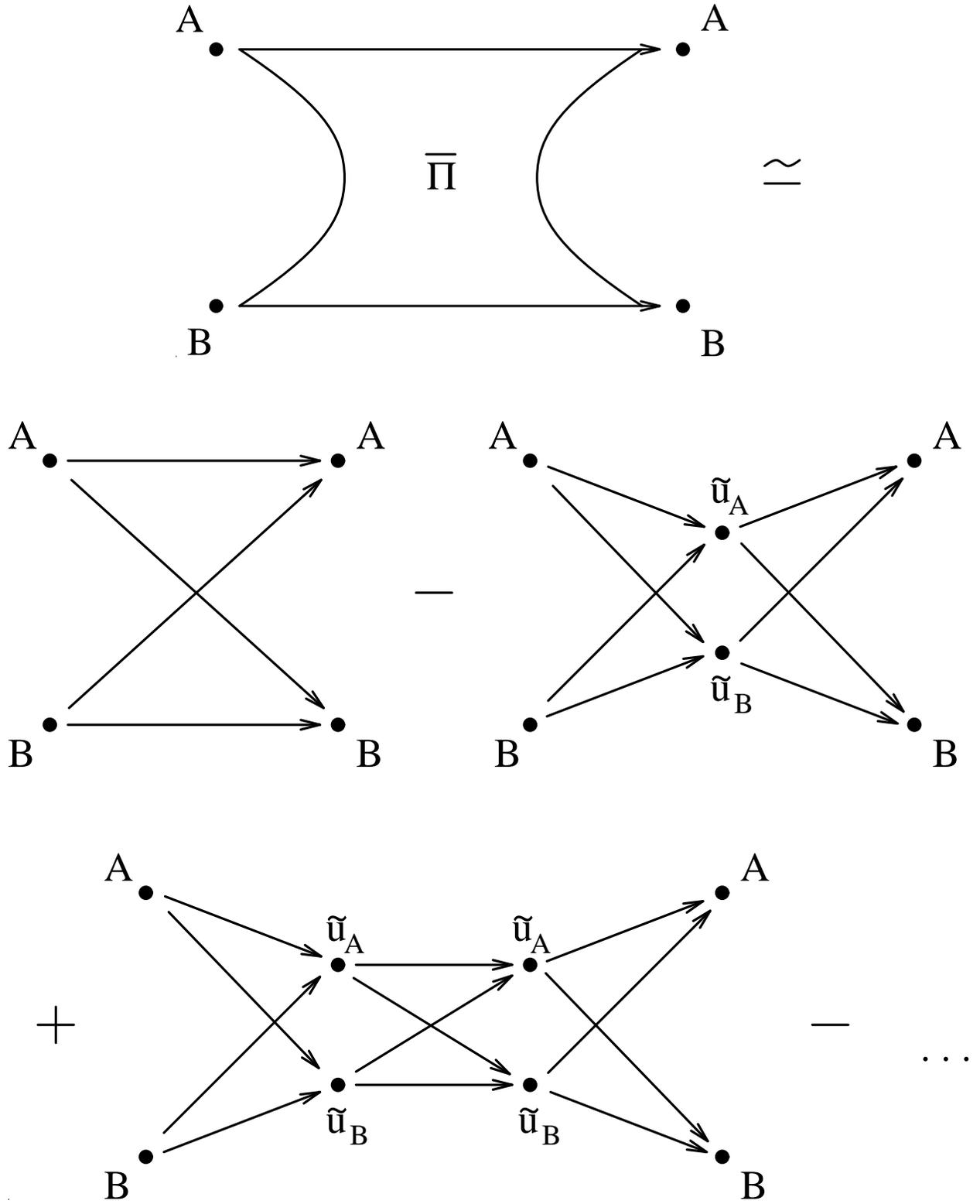
\begin{figure}[tbp]
\caption{Diagrammatic representation of processes to second order
included in the
weighted expansion of the excitonic polarizability. The electron hole
pair can be excited on an $A-$ or a $B$-site and subsequently be
 scattered on either of such sites with the
effective contact
potential strengths  $\tilde u_A$ and $\tilde u_B$, respectively.
 Every
connecting line between two sites denotes a factor of a
non-interacting
weighted polarizablity $\bar \Pi_0^{XY}$ ($X,Y \in
\{A,B \}$), where $X$ and $Y$ are chosen to match the
 type of the start and the end site of an
arrow and all conceivable paths occurring between the excitation
and the recombination of the electron-hole pair are being summed
 over.}
\label{Fig1}
\end{figure}

We introduce the following attractive weighted contact scattering
interaction strengths
\begin{equation}
\tilde u_A= \frac {u_A} c \hspace{1mm} ; \hspace{20mm} \tilde u_B=
 \frac {u_B}
{1-c} \hspace{1mm}, \label{eb0}
\end{equation}
associated with the $A-$ and $B$-components of the medium, whereby
 $u_A$
and $u_B$ represent bare scattering strengths, and the limit of
 uniform
interaction strengths corresponds to $u_A \rightarrow u_B
\rightarrow u$.
If one distinguishes now the different types of sites involved in the
propagation of the two particles during a scattering process, the
expansion of the scattering series can be represented pictorially
as in Fig. \ref{Fig1}.
Mathematically this corresponds to writing
\begin{eqnarray}
\bar \Pi & \simeq & \bar \Pi_0^A + \bar \Pi_0^B - \left[ \bar \Pi_0^A
 \tilde u_A \bar \Pi_0^A  +
\bar \Pi_0^B \tilde u_B \bar \Pi_0^B  \right] \nonumber \\
& + & \left[ \bar \Pi_0^A \tilde u_A \bar \Pi_0^{AA} \tilde
u_A \bar \Pi_0^A +
\bar \Pi_0^A \tilde u_A \bar \Pi_0^{AB} \tilde u_B
\bar \Pi_0^B  \right. \nonumber \\
& + & \left. \bar \Pi_0^B \tilde u_B \bar \Pi_0^{AB} \tilde u_A
\bar \Pi_0^A  +
\bar \Pi_0^B \tilde u_B \bar \Pi_0^{BB}  \tilde u_B
\bar \Pi_0^B \right]  - \cdots ,\label{eb1}
\end{eqnarray}
where both, the diagrams in Fig. \ref{Fig1} and the terms written
 down in
(\ref{eb1})  above represent the expansion to second order.
 These and higher order terms can can be conveniently rewritten in
 a $2 \times 2$ matrix
scattering formalism by introducing

\def\pio%
{\begin{array}{cc}
\bar \Pi^{AA}_0 & \bar \Pi^{AB}_0 \\
\bar \Pi^{AB}_0 & \bar \Pi^{BB}_0
\end{array}}
\def\uu%
{\begin{array}{cc}
- \tilde u^A & 0 \\
0 & - \tilde u^B
\end{array}}
\def\vect%
{\begin{array}{c}
\bar \Pi^A_0 \\
\bar \Pi^B_0
\end{array}}

\begin{equation}
\begin{array}{ccccc}
 \vec \Pi_0 = \left( \vect \right), & \hspace{5mm} & {\bf \hat U} =
\left( \uu \right), & \hspace{5mm} & {\bf \hat \Pi}_0  = \left( \pio
 \right).
\end{array}
\end{equation}
Equation (\ref{eb1}) then goes over to
\begin{equation}
\bar \Pi = \bar \Pi_0 +  \vec \Pi_0^T {\bf \hat U} \vec \Pi_0 +
\vec \Pi_0^T {\bf \hat U \hat \Pi}_0 {\bf \hat U } \vec \Pi_0 + \cdots
 \label{eb2}.
\end{equation}
The matrix products can be summed to give
\begin{equation}
\bar \Pi = \bar \Pi_0 + \vec \Pi_0^T {\bf \hat U} \sum_{n=0}^{\infty}
 \left[ {\bf \hat \Pi}_0 {\bf \hat U} \right]^n
 \vec \Pi_0 \label{eb3},
\end{equation}
which can be evaluated as a matrix geometric series as
\begin{equation}
\sum_{n=0}^{\infty} \left[ {\bf \hat \Pi}_0 {\bf \hat U} \right]^n =
\left[ {\bf 1 - \hat \Pi}_0 {\bf \hat U} \right]^{-1},
\end{equation}
and equation (\ref{eb3}) is calculated to yield
\begin{equation}
\bar \Pi = \bar \Pi_0^A +  \bar \Pi_0^B -
\frac { \tilde u_A ( \bar \Pi_0^A )^2 (1+
\tilde u_B \bar \Pi_0^{BB}  ) + \tilde u_B
( \bar \Pi_0^B )^2 (1+ \tilde u_A \bar \Pi_0^{AA} ) -2 \bar
\Pi_0^A \tilde u_A \bar \Pi_0^{AB} \tilde u_B
\bar \Pi_0^B } {(1+ \tilde u_A \bar \Pi_0^{AA})(1+ \tilde
u_B \bar \Pi_0^{BB}  ) - \tilde u_A \tilde u_B (\bar \Pi_0^{AB}
)^2} \label{eb4}.
\end{equation}
Equation (\ref{eb4}) is the central result of this work and much of
the further discussion will be based on its properties.

\subsection{Limiting Behavior}

In order to test  the usefulness of equation (\ref{eb4}) it is
necessary to investigate its behavior in various limits of disorder.

\subsubsection{Weak disorder limit; uniform interaction}

To establish a connection with formula (\ref{ea2_2})
we first study the limit of low disorder and asymptotically equal
bare contact interaction strengths $u^A = u^B = u$. It is well known
 \cite{bi2}
that in this limit the CPA self energy goes over to the virtual
 crystal
limit $\Sigma_{c/v} \rightarrow cV_{c/v}$ and the vertex correction
 to $ \Lambda
\rightarrow c(1-c)V_cV_v$. Accordingly, the weights in the two
 particle function defined in
(\ref{mb7_1})--(\ref{mb7_3})
go over to $\xi \rightarrow c$, $\eta \rightarrow (1-c)$
and $\gamma \rightarrow c(1-c)F_cF_v$. As a result of this it is
possible to pull the two weights $\xi$ and $\eta$ multiplied with the
unweighted two particle function out of the energy convolution, since
they are now independent of energy. The impurity weighted versions
of the polarizability can therefore be expressed in terms of the
 unweighted function as $ \bar \Pi^A = c  \bar \Pi$ and $
\bar \Pi^{AA} = c^2 \bar \Pi  + c(1-c) \Omega$, where
$\Omega(i \omega) \equiv  - \beta^{-1} \sum_{iz} F_c(i \omega)
 F_v(iz- i \omega)$. The behavior of the host weighted functions
 and the mixed one follow in
analogy. Through inserting into (\ref{eb4}) it is found that the
 scattering term factorizes as
\begin{equation}
\bar \Pi  \rightarrow \bar \Pi_0  - \frac {\bar
\Pi_0  u \bar \Pi_0  (1+ u \Omega)} {(1+u \bar
\Pi_0 )(1+ u \Omega)} \label{eb5},
\end{equation}
which is identical to (\ref{ea2_2}).

\subsubsection{Strong disorder or split band limit}

In Ref. \onlinecite{bi12} we had shown explicitly that in the regime
 of strong
disorder, when the widths of the single particle bands become
negligible compared to the disorder strengths involved, the mixed
 component
of the CPA two particle function $K^{AB}$ go to zero whereas the
double weighted functions $K^{AA/BB}$ become effectively identical
 to the
single weighted ones $K^{A/B}$. The same behavior also translates to
the weighted polarizabilities and therefore in this limit (\ref{eb4})
 goes over to
\begin{equation}
\bar \Pi  = \frac {\bar \Pi_0^A } {1+ \tilde u^A
\bar \Pi_0^A } + \frac {\bar \Pi_0^B } {1+ \tilde u^B
\bar \Pi_0^B }. \label{eb6}
\end{equation}
Even though this already looks very similar to (\ref{ea5}) it is
not trivially the same. However we showed in Ref.
\onlinecite{bi12} that in this limit
\begin{equation}
 K^{A/B} \rightarrow x^{A/B} \sum_k G^{A/B}_{c \hspace{2mm} crys}(k)
G^{A/B}_{v \hspace{2mm} crys}(-k)= x^{A/B} K^{A/B}_{crys} \label{eb7}
\end{equation}
where $x^{A/B}$ is the concentration $c$ and $(1-c)$ of the $A-$ and
 $B$-component, respectively, and the subscript $_{crys}$ is chosen
 to label
the corresponding two particle functions
of the pure $A-$ and $B$-media. This in connection with the
definition of $\tilde
u^A$ and $\tilde u^B$ from (\ref{eb0}) finally shows that
(\ref{ea5}) and (\ref{eb4}) are indeed
identical in this limit.

\subsubsection{General behavior}

In order to examine the predictions that (\ref{eb4}) makes for
a general case, we introduce the following
 definitions
\begin{equation}
\begin{array}{ccc}
\bar \Pi_0^A  \equiv \chi \bar \Pi_0;  &\hspace{15mm} & \bar
\Pi_0^{AA} \equiv \chi \zeta \bar \Pi_0   + \Gamma , \\
\end{array} \label{eb8}
\end{equation}
where $\Gamma(i \omega) \equiv  - \beta^{-1} \sum_{iz}
 \gamma(iz,i \omega -iz)$.
Due to probability conservation the other weighted
functions are defined as
\begin{equation}
\begin{array}{ccccc}
\bar \Pi_0^B  = (1-\chi) \bar \Pi_0 ; & \vspace{5mm} & \bar
\Pi_0^{AB}  = \chi (1-\zeta) \bar \Pi_0  - \Gamma; & \vspace{5mm}
 & \bar
\Pi_0^{BB}  = (1- 2 \chi + \chi \zeta) \bar \Pi_0 +
\Gamma .\\
\end{array} \label{eb9}
\end{equation}
In a general case, the polarization weights $\chi$ and $ \zeta $ are
only very indirectly
related to the original weights $ \xi $ and $ \eta $ for the
 unintegrated two particle
function $K$. They may assume different values for any particular
 energy
at which the polarizability is considered and therefore they
 generally depend on energy.
Inserting this into (\ref{eb4}) we find that
\begin{eqnarray}
& \bar \Pi & = \bar \Pi_0 - \bar \Pi_0^2
 \left\{ \left[ \chi^2 \tilde u^A + (1- \chi)^2 \tilde u^B \right] +
 \tilde
u^A \tilde u^B \left[ \chi ( \zeta - \chi) \bar \Pi_0 + \Gamma \right]
\right\}  \nonumber \\
 & \times & \left\{ 1+ \left[ \chi \zeta \tilde u^A + (1 -2 \chi
+ \chi \zeta) \tilde u^B +
\tilde u^A \tilde u^B \Gamma \right]
\bar \Pi_0 + (\tilde u^A + \tilde u^B) \Gamma + \tilde u^A \tilde u^B
\chi ( \zeta  -\chi) \bar \Pi_0^2 \right\}^{-1}. \label{eb10}
\end{eqnarray}
If this this is compared to a corresponding Dyson equation for the
polarizability with a renormalized self-energy-like expression M
\begin{equation}
\bar \Pi = \bar \Pi_0 + \bar \Pi_0 {\rm M} \bar \Pi  \label{eb11},
\end{equation}
this expression M, which can also be viewed as a disorder average
``dressed'' interaction, is found to be
\begin{equation}
{\rm M}  = - \frac { \left[ \tilde u^A \chi^2 + (1- \chi)^2 \tilde u^B
\right] + \tilde u^A \tilde u^B \chi (1 - \chi) \Phi }
{1+ \left[ \tilde u^A + \tilde u^B \right] \chi (1 - \chi) \Phi}
\label{eb12},
\end{equation}
where we had introduced $\Phi = \left[ \chi ( \zeta -\chi) \bar
\Pi_0 + \Gamma \right] / \left[ \chi (1 - \chi)\right] $.
Introducing a generalized  energy dependent  average interaction
strength
\begin{equation}
\bar u = \chi^2 \tilde u^A + (1- \chi)^2 \tilde u^B \label{eb13},
\end{equation}
and a generalized contact potential fluctuation strength
\begin{equation}
\Delta = \chi \tilde u^A - (1- \chi) \tilde u^B \label{eb14},
\end{equation}
equation (\ref{eb12}) can be rewritten as
\begin{equation}
{\rm M} = - \bar u + \frac {\chi(1-\chi) \Delta^2 \Phi }
{1 + \left[ \bar u + \Delta (1 - 2 \chi) \right] \Phi } \label{eb15}.
\end{equation}
Finally introducing the renormalized function $\Psi = \Phi / (1
+ \bar u \Phi )$ one obtains
\begin{equation}
{\rm M} = - \bar u + \frac { \chi (1-\chi ) \Delta^2 \Psi }
{ 1 + \Delta (1 - 2 \chi )  \Psi } \label{eb16}.
\end{equation}
We argue that this result represents a generalized form of an average
T-matrix approximation (ATA) of the polarizability of the
medium with respect to the disordered contact
potential relative to its ``mean'' background-value $\bar u$ which
 is taken as the
basis of an energy dependent background medium similar to a virtual
 crystal
(VCA) energy. To motivate our comparison more, we recapitulate the
 main features of
 the ATA for a single
particle  propagator $G$ in a
binary disordered medium with on-site potentials $- \varepsilon^A$ and
$- \varepsilon^B$ on the $A-$ and $B$-components, respectively. The
contact potentials have been defined with a negative sign here to
match the definition of $u^A$ and $u^B$ from before, which were
introduced as attractive interactions on an equivalent footing.
The averaged T-matrix equation for $G$ can be written as
\begin{equation}
\bar G = \bar G_0 + \bar G_0 \bar T \bar G_0 \label{eb17},
\end{equation}
in which $\bar G_0$ is the virtual crystal propagator
\begin{equation}
\bar G_0 = \frac g {1 + \bar \varepsilon g} \label{eb18},
\end{equation}
where we have set $\bar \varepsilon = c \varepsilon^A + (1-c)
\varepsilon^B$ and  $g$ is the the propagator for a  medium without
the on-site potentials. The
average T-matrix $\bar T$ is defined as
\begin{equation}
\bar T = \frac {- c [\varepsilon^A - \bar \varepsilon]} {1 + [
\varepsilon^A - \bar \varepsilon ] F_0} + \frac
{-(1-c) [ \varepsilon^B - \bar \varepsilon] } {1 + [ \varepsilon^B -
 \bar \varepsilon ] F_0} \label{eb19},
\end{equation}
where $F_0= \bar G_0(m,m)$. Introducing furthermore  $\delta =
\varepsilon^A - \varepsilon^B$, the corresponding ATA selfenergy
$\Sigma_{ATA}$ relating $g$ and $\bar G$ can be written as
\begin{equation}
\Sigma_{ATA} = -\bar \varepsilon + \frac {\bar T} {1+ \bar T F_0} =
 -\bar \varepsilon +\frac {c(1-c) \delta^2 F_0} {1+ (1-2c)
\delta F_0} \label{eb20}.
\end{equation}
The aforementioned analogy hence builds on the correspondence of
 the quantities
\begin{equation}
\begin{array}{ccccccc}
\bar \varepsilon & \longleftrightarrow & \bar u & \hspace{15mm} &
 \delta & \longleftrightarrow & \Delta \\
g & \longleftrightarrow & \bar \Pi_0 & \hspace{15mm} &
\Sigma_{ATA} & \longleftrightarrow & {\rm M} \\
\bar G  & \longleftrightarrow & \bar \Pi &
\hspace{15mm} & F_0  & \longleftrightarrow & \Psi \\
\end{array} \label{eb21}
\end{equation}
This correspondence becomes exact in the asymptotic limits discussed
before and continues to hold qualitatively in an intermediate regime.
The difficulty in finding a rigorous comparison in the
most general case stems again from difficulty of finding general
expressions for the energy dependence of
the integral weights $\chi$ and  $ \zeta $ and the integrated diagonal
 correction term $\Gamma$.

The reason why our weighted expansion scattering expansion can be
expected to yield a better result than an undifferentiated decoupling
of the Dyson equation for the contact potential, as it is shown to do
 by our
numerical calculations in the next section, can be understood from
the argument that averages over higher moments of $\Pi_0$ which occur
 in this
expansion such as $\langle \Pi_0^2 \rangle$ and $\langle \Pi_0^3
\rangle$ are better approximated by a
decoupling as
$\langle \Pi_0^2 \rangle \simeq ( \bar \Pi_0^A )^2/c + ( \bar
\Pi_0^B )^2/(1-c)$ and $\langle \Pi_0^3 \rangle \simeq
( \bar \Pi_0^A )^2 \bar \Pi_0^{AA} /c^2 + 2 \bar \Pi_0^A \bar \Pi_0^B
\bar \Pi_0^{AB}/c(1-c) + \bar \Pi_0^{BB} ( \bar \Pi_0^B )^2/(1-c)^2$
 rather than $\langle
\Pi_0^2 \rangle \simeq \bar \Pi_0^2$ and  $\langle
\Pi_0^3 \rangle \simeq \bar \Pi_0^3$, respectively, because the former
approximation reduces a wrong weighting and overcounting of scattering
 processes that the
latter approximation erroneously infers.

\section{Numerical results}

In this section we discuss a numerical implementation of our previous
results within a commonly used model in three dimensional systems.
Since within the quasi-ATA contact
 potential model of the electron-hole interaction the excitonic
polarizability can be obtained directly by inserting the weighted
components of the average polarizability for the free particle system
 into equation (\ref{eb4}), we draw strongly from the results
 obtained for the non-interacting system in Ref. \onlinecite{bi12}.

The main input necessary for the numerical implementation of a single
site CPA are the single particle densities of states for the
 conduction and valence band. As
in Ref. \onlinecite{bi12} we take
\begin{equation}
\begin{tabular}{ c c }
$ \rho_{\mu} (E) = \displaystyle{ \frac 2 {\pi w_{\mu}^2}}
\sqrt{w_{\mu}^2 - E^2} $ &
$ \mid E \mid \leq w_{\mu} $ \\
\\
$ \rho_{\mu} (E) = 0 $ & $ \mid E \mid \leq w_{\mu} $ \label{n1}.
\end{tabular}
\end{equation}
It is well known that this density of states does not exhibit any of
the finer structure  of a real system, but it includes the more global
features of a wide class of systems giving a finite band width and
the appropriate van Hove singularities at the band
edges. Moreover, it matches well with the spirit of a single
site CPA which, even though it reproduces the global influence of
disorder on the system  correctly in all regimes, cannot account for
more
detailed structure brought about by the scattering of particles
off clusters of impurities which usually involve momentum dependent
self energy contributions whose evaluation would require a more
 specific
knowledge of the single particle dispersion laws.

For amalgamation type solid solutions,
i.e. alloys whose densities of
states of the pure $A-$ and $B$-substances overlap to a great extent
 in the
respective conduction and valence bands, CPA results  already exist
\cite{bi2} for the case $u_A=u_B=u$. We shall therefore focus on the
 region of parameters which
corresponds to intermediate disorder, where
our approximation provides significantly better results than earlier
approaches. A restriction to a rather
specific region of parameter space becomes necessary since we now
have six independent parameters which govern the behavior of our
spectra, i.e. the concentration $c$, the relative valence
-- conduction band
width $w_v/w_c$, the disorder strengths $V_c$ and $V_v$ in the
 respective bands and the
corresponding disordered contact interaction strengths $u_A$ and
 $u_B$, accounting for
the dynamic part of the electron-hole correlations.

In determining the region of interest we note that, as pointed out
 by many workers (cf. for example the
review article by Rashba \cite{bia6}), the CPA produces a ``pseudo''
 gap in the
single particle DOS at all concentrations once the disorder strengths
 exceed the half-width
 of the single particle
bands involved, $V_{\mu} \geq w_{\mu}$, and also translates
to the joint DOS in some regimes. In more realistic systems,
however, usually a true gap
only exists once the disorder strengths exceed the full band widths
$ V_{\mu} \geq
2 w_{\mu}$,
even though in the regime $ 2 w_{\mu} \geq V_{\mu} \geq
w_{\mu}$ the number of states in the region of the pseudo gap is
 strongly
suppressed. Because of this, we shall concentrate our attention to
 the beginning
region of the true gap behavior and choose $ \mid V_{\mu} \mid = 3
w_{\mu}$.

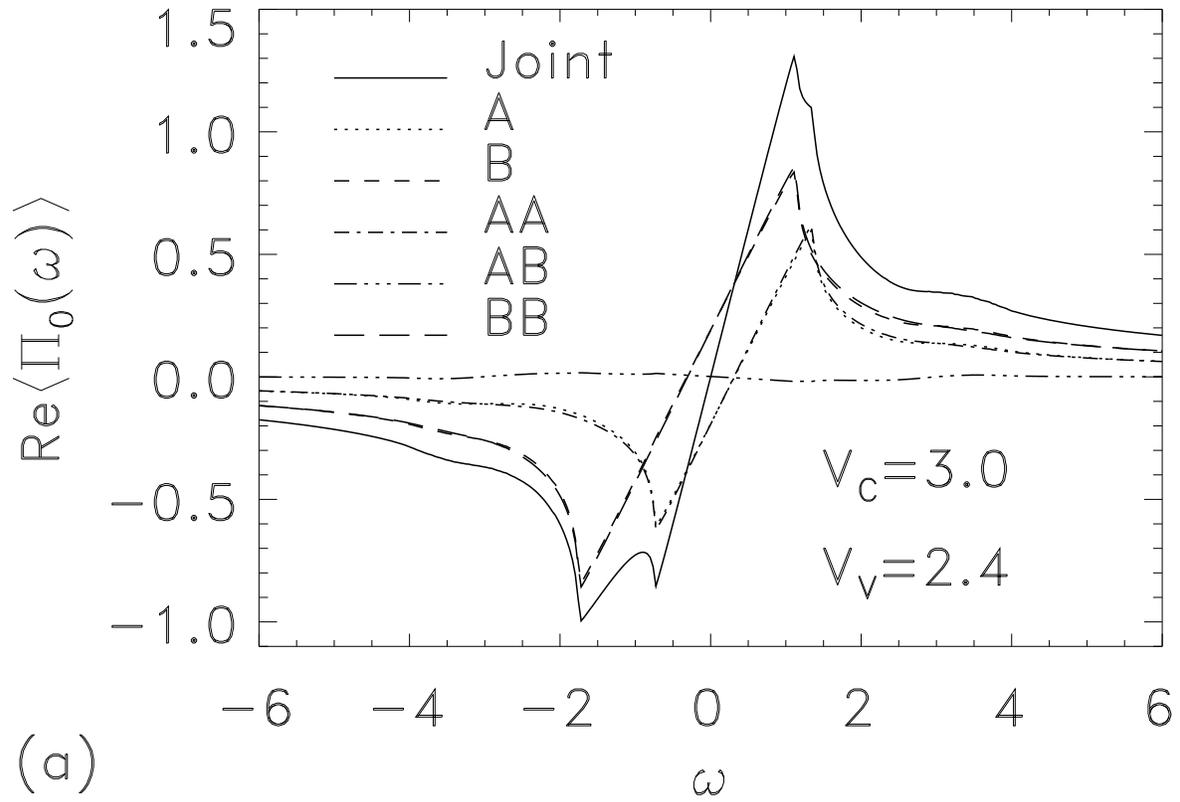
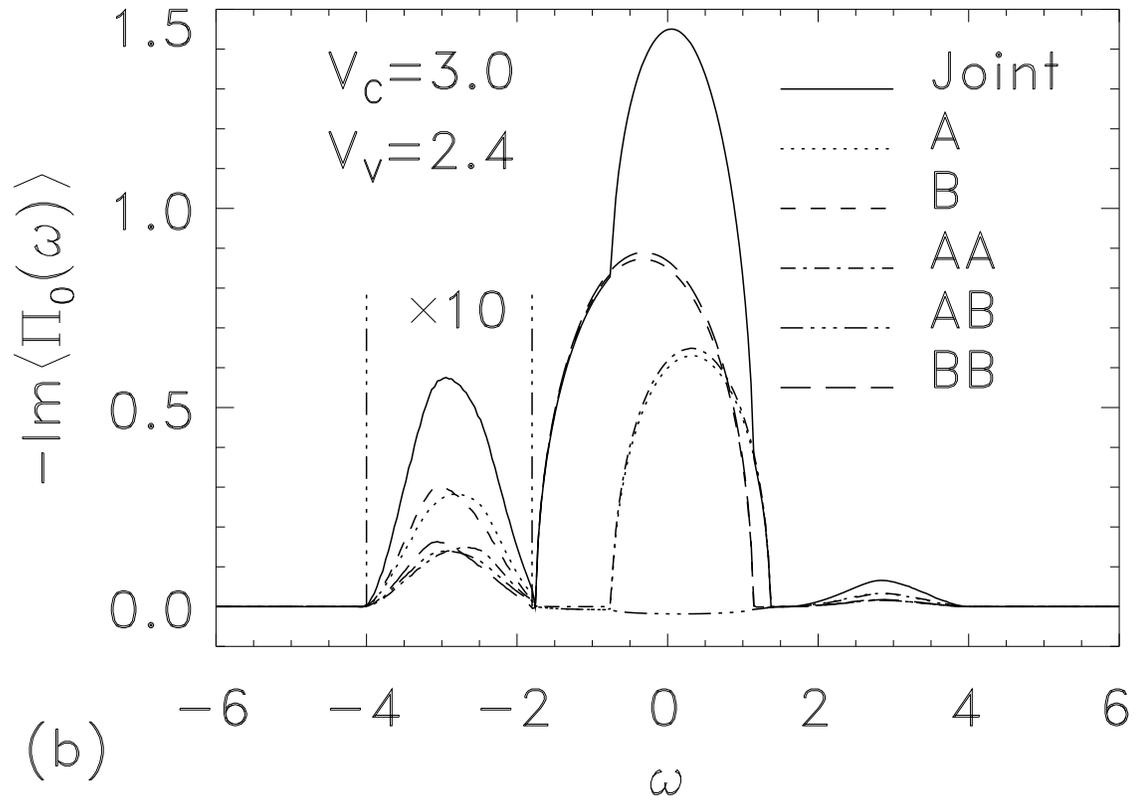
\begin{figure}[tbp]
\caption{(a) Real and (b) negative imaginary parts of the
non-interacting polarizability
 for parallel disorder at c=0.35. All
energies scale with the conduction half-band width $w_c$ which has
 been set
to unity, $w_c \equiv 1$, while $w_v$ is taken to be $w_v =
0.8$. Since this
 spectrum is taken at
disorder strengths $V_c= 3.0$ and  $V_v= 2.4$, i.e. at the
beginning of the true gap region of the CPA, the single and double
weighted functions coincide to a great extent, and two flanks of the
spectra which to a substantial degree contain mixed $AB$-components
 have
split off sideways (cf. the lower end of the joint
DOS in (b) where the values were enlarged by a factor of 10).
 However, it becomes clear from the spectra in
Fig. 3 that the residual interference of $A-$ and $B$-absorption
amounts to a significant influence on the excitonic absorption as the
corresponding electron-hole interactions are introduced.}
\label{Fig2}
\end{figure}

In calculating the absorption spectra, we have first obtained the
 weighted
components of the polarizability without electron-hole
interactions. Figures \ref{Fig2} (a), (b) and \ref{Fig4} (a), (b) show
 the real and negative
imaginary part of $\bar \Pi_0$ and all its decompositions into single
 and
double weighted components for parallel disorder $sgn(V_c)=sgn(V_v)$
 and
anti-parallel disorder $sgn(V_c)=- sgn(V_v)$,
respectively. Specifically,
 the conduction and valence band
disorder have been chosen to be $ V_c = 3.0$ and $V_v = \pm 2.4$ in
 connection with
the half-bandwidths $w_c =1.0$ and $w_v=0.8$.
It should be noted, that all energies occurring
in our results effectively scale with the half-width of the
 conduction band
$w_c$ as it is normalized to unity.

We can see from Fig. \ref{Fig2} (b) and Fig. \ref{Fig4} (b), that
 in the region of
parameters considered, the joint density of states has already split
 quite
clearly into distinct $A-$ and $B$-parts for both cases of disorder
and the difference of the
double and single weighted components has become relatively small due
to the dominance of the total diagonal matrix elements of the two
particle function in this regime. Note that in all the figures for the
non-interacting polarizability only the single $AB$-function has been
 plotted and the
components of the spectra sum as $\bar \Pi^{A/B}_0 = \bar
\Pi^{AA/BB}_0 +
\bar \Pi^{AB}_0$ and $\bar \Pi_0 = \bar \Pi^{AA}_0 + \bar \Pi^{BB}_0
 + 2
\bar \Pi^{AB}_0$.

In the case of parallel disorder of Fig. \ref{Fig2} (b), the joint
 DOS is divided up into
into a central bulk part which is mainly constituted from $AA-$ and
$BB$-transitions and a separate set of flanks on either side of it,
which contain a notable amount of $AB-$transitions. In the case of
 anti-parallel disorder of Fig. \ref{Fig4} (b), on the other hand, it
 decomposes into two separate
parts with a pronounced gap between them, whereby the upper part
consists mainly of $AA-$ and
the lower part of $BB$-transitions.
Even though the joint function appears to be very close to zero in the
gap region for this case, we discover a finite contribution of
double weighted components centered between the split bands
\cite{bi12}. These
states can be considered ``mute'' in the absence of any electron-hole
 interaction since cumulatively they do not contribute to the
polarization.

It appears from these results as if the system is already very close
to the strong disorder asymptotic behavior where
the split components are essentially independent, but we shall find
that this
is indeed not the case, agreeing with the result
of Onodera and Toyozawa \cite{bia2} who state that this regime is
reached only beyond a relative disorder strength of about
$V_{\mu}/w_{\mu} = 10$. If in the interacting case states are
 pulled down from
the upper band into this region by means of the electron-hole
interaction, they will experience a significant
broadening and hence change the excitonic absorption
in a large region as as will be apparent from the plots in Fig.
\ref{Fig5}.

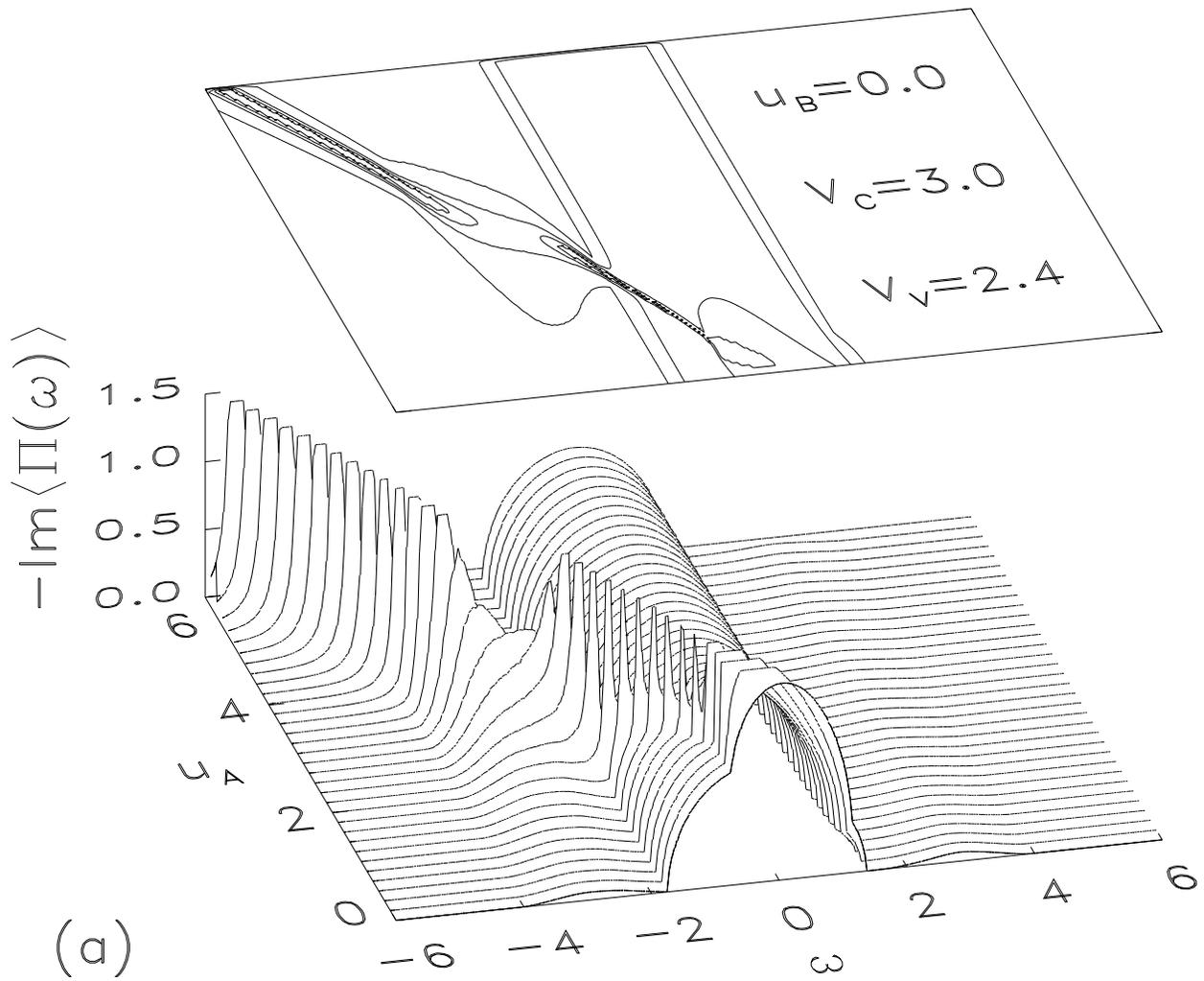
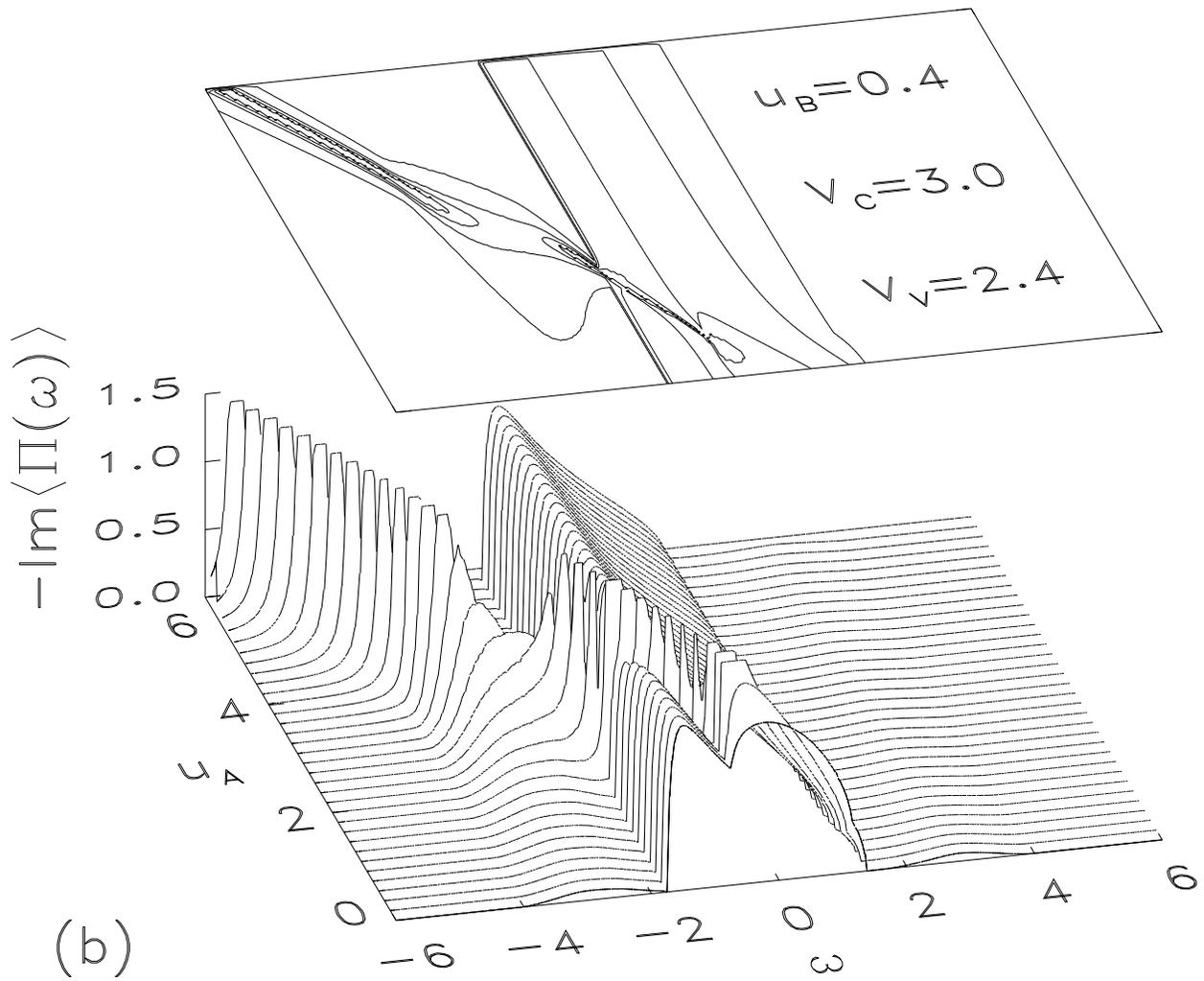
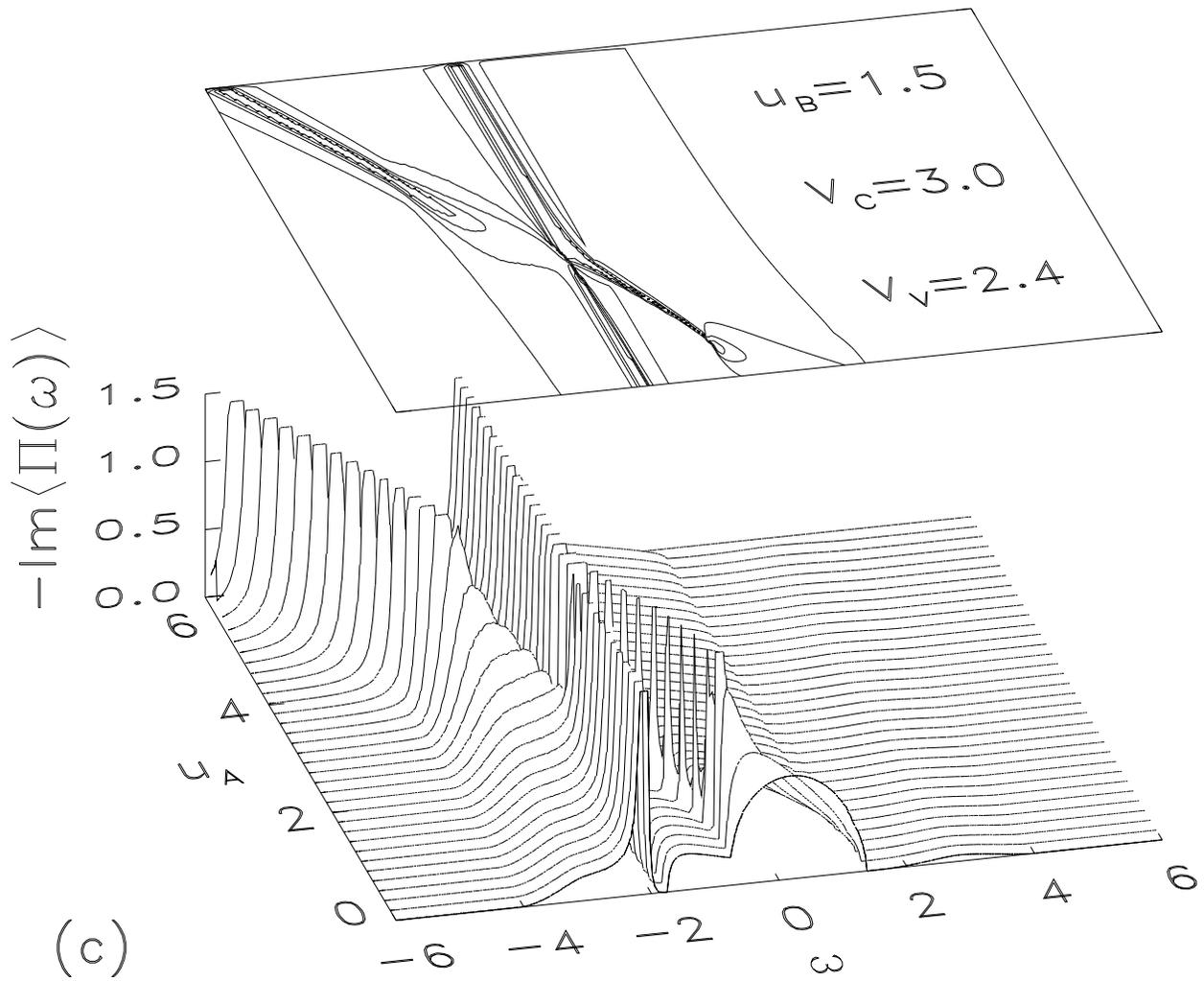
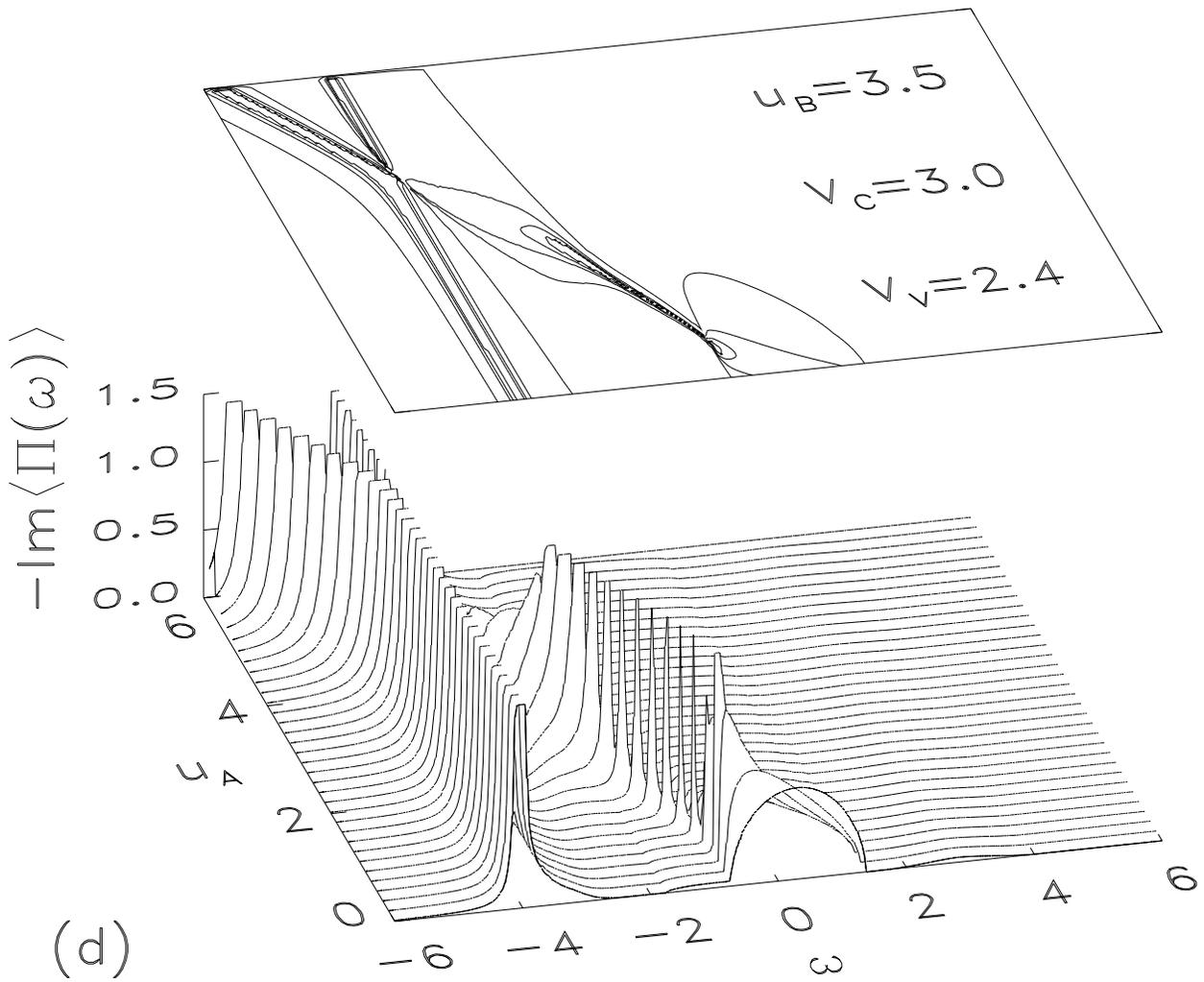
\begin{figure}[tbp]
\caption{Excitonic absorption spectra created by means of equation
 (34)
from the weighted components of the non-interacting
spectrum in Fig. 2. In each of the plots (a), (b), (c), (d) the
 contact
 potential strength $u_B$ associated with $B$-atoms is kept at a
constant value while $u_A$ is increased from 0 to 6 from the front to
the back of the set of overlayed plots. The overlayed contour plots
 show
the paths and broadenings of the excitons as the parameter strengths
 are
varied. The contour lines are taken at level heights of
0.1, 0.5, 1.0, 1.5, respectively. Comparing the plots (a)
$u_B=0.0$, (b) $u_B=0.4$, (c) $u_B=1.5$, (d) $u_B=3.5$. one can see
that mutual interference effects of both interactions are relatively
small.
 The main such interference effect is that $B$-resonance attains
some broadening once the $A$-resonance has passed it, cf. Fig. 3 (c).}
\label{Fig3}
\end{figure}

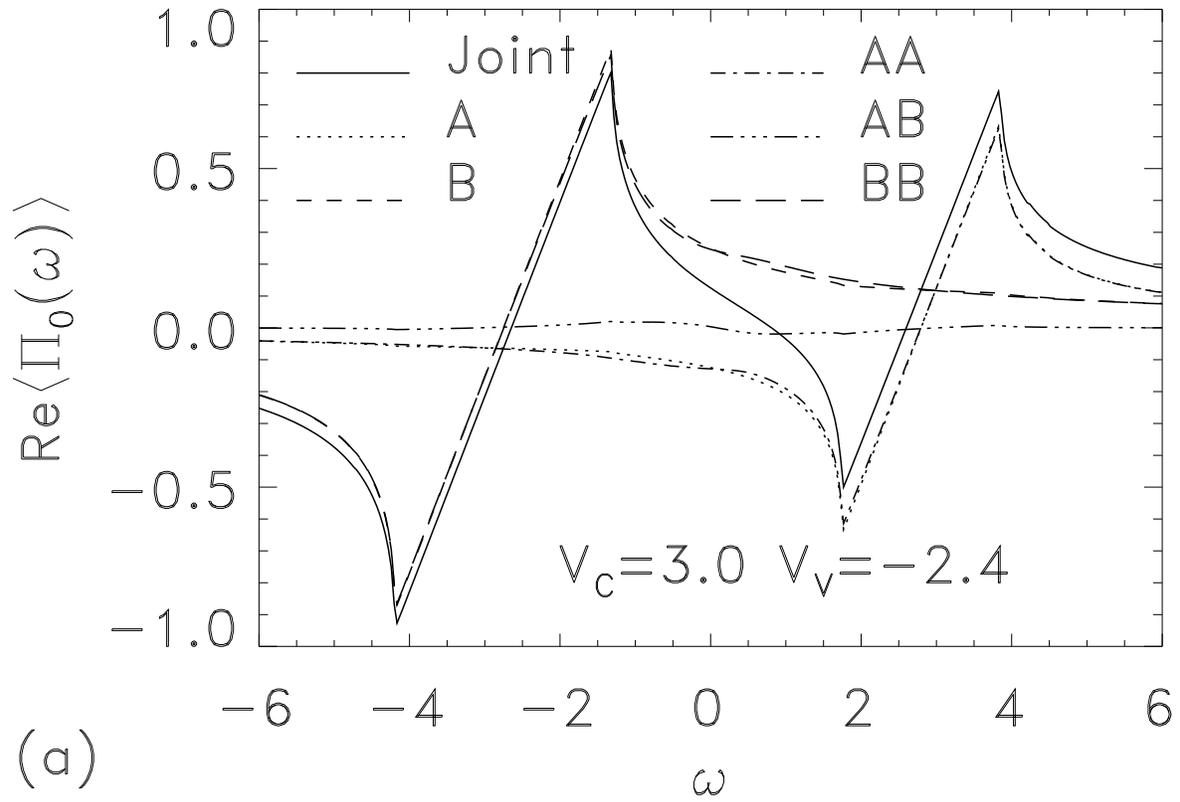
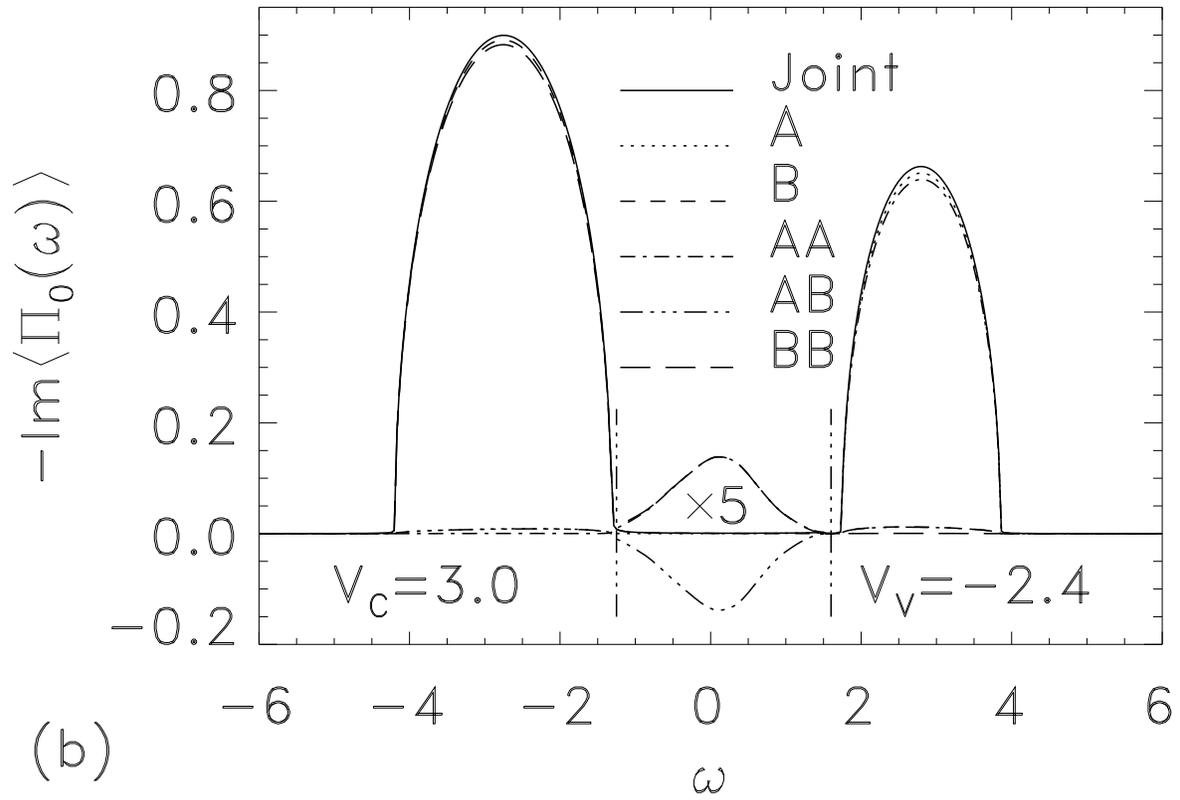
\begin{figure}[tbp]
\caption{(a) Real and (b) negative imaginary parts of the
non-interacting
 polarizability
 for anti-parallel disorder and parameters $c=0.35$, $V_c= 3.0$,
 $V_v= -2.4$,
$w_c = 1.0$ and $w_v = 0.8$. The anti-parallel direction of the band
offsets causes the imaginary part (joint DOS) in (b) to split into two
 distinct components
mainly constituted by $A-$ and $B$-transitions, respectively,
 separated
by a wide gap. Even though the cumulative function is very close to
zero in the central region, the double weighted components are
finite there (cf. the central region in (b) where the values were
enlarged by a factor of 5) and hence they lead to a
substantial broadening if  an excitonic
resonance is pulled out of the higher lying $A$-component into this
region, as can be seen from the excitonic spectra in Fig. 5. }
\label{Fig4}
\end{figure}

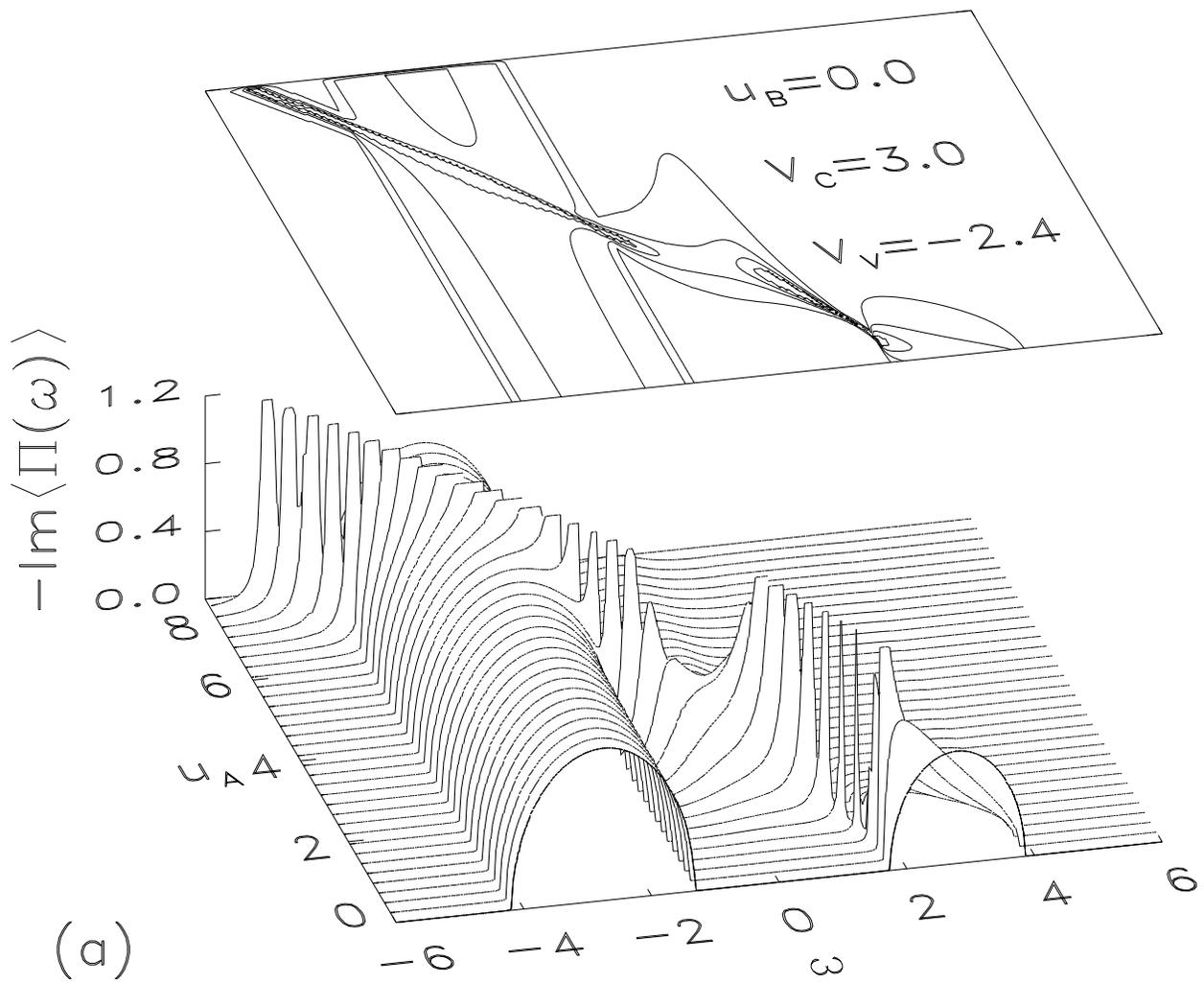
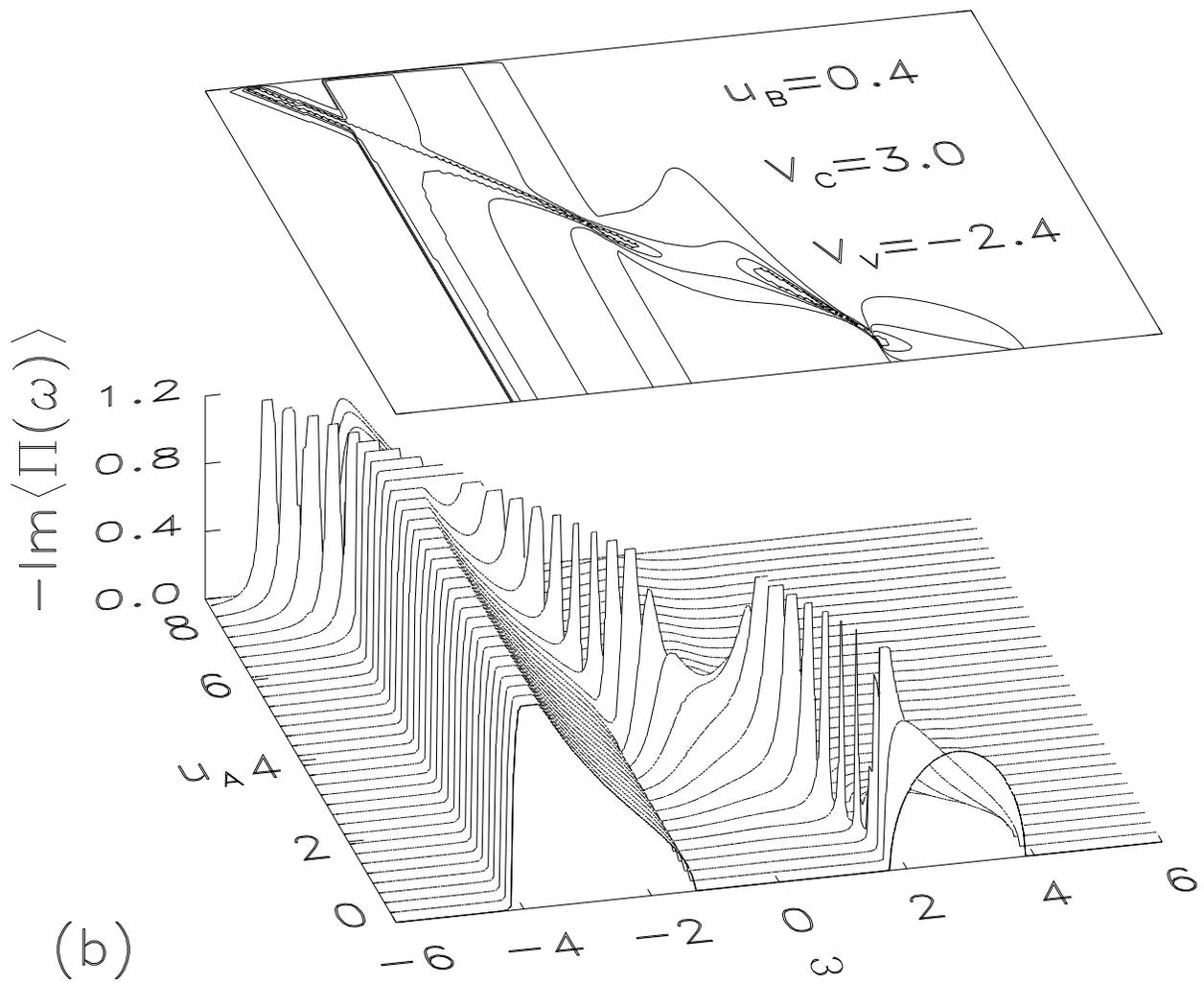
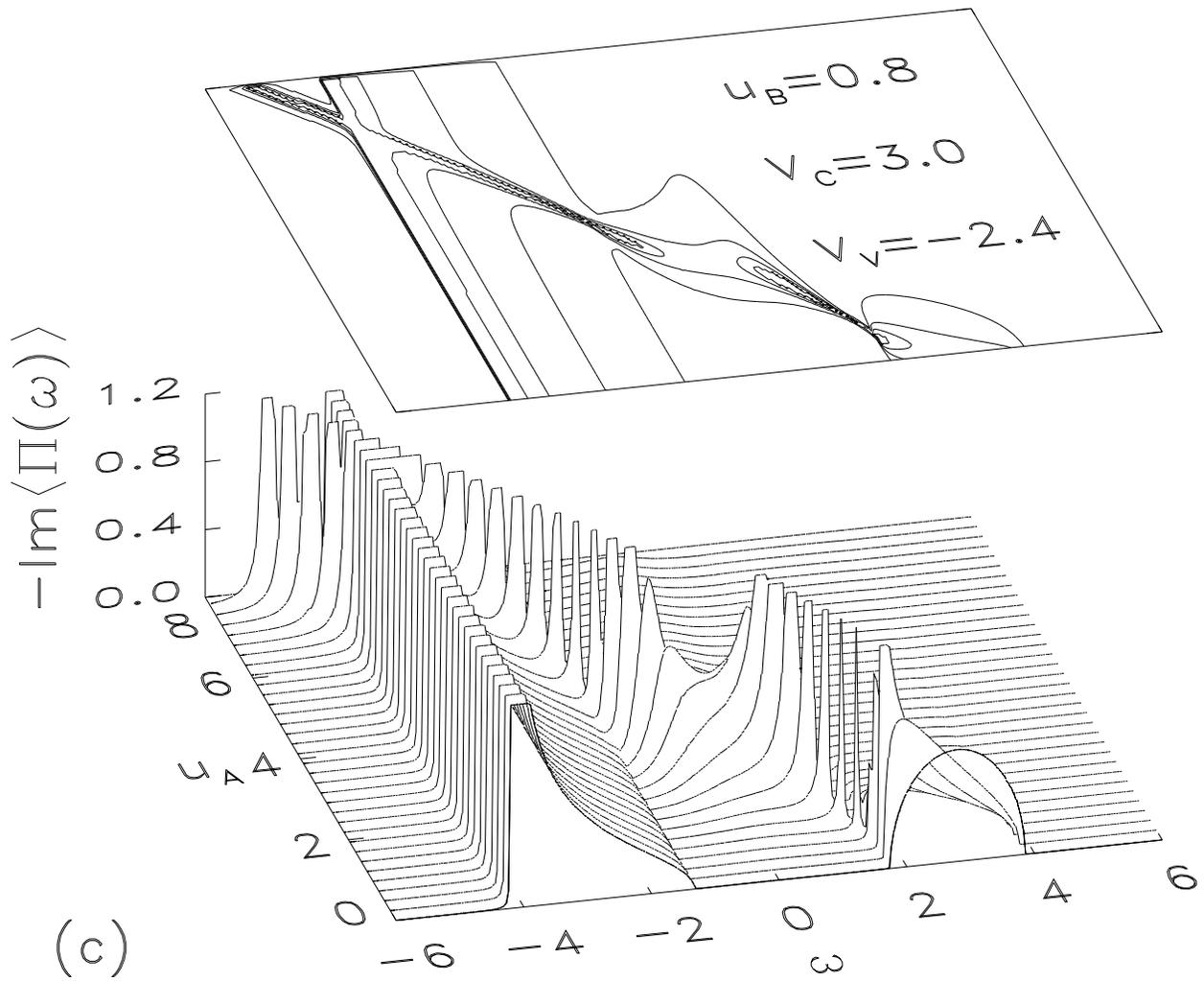
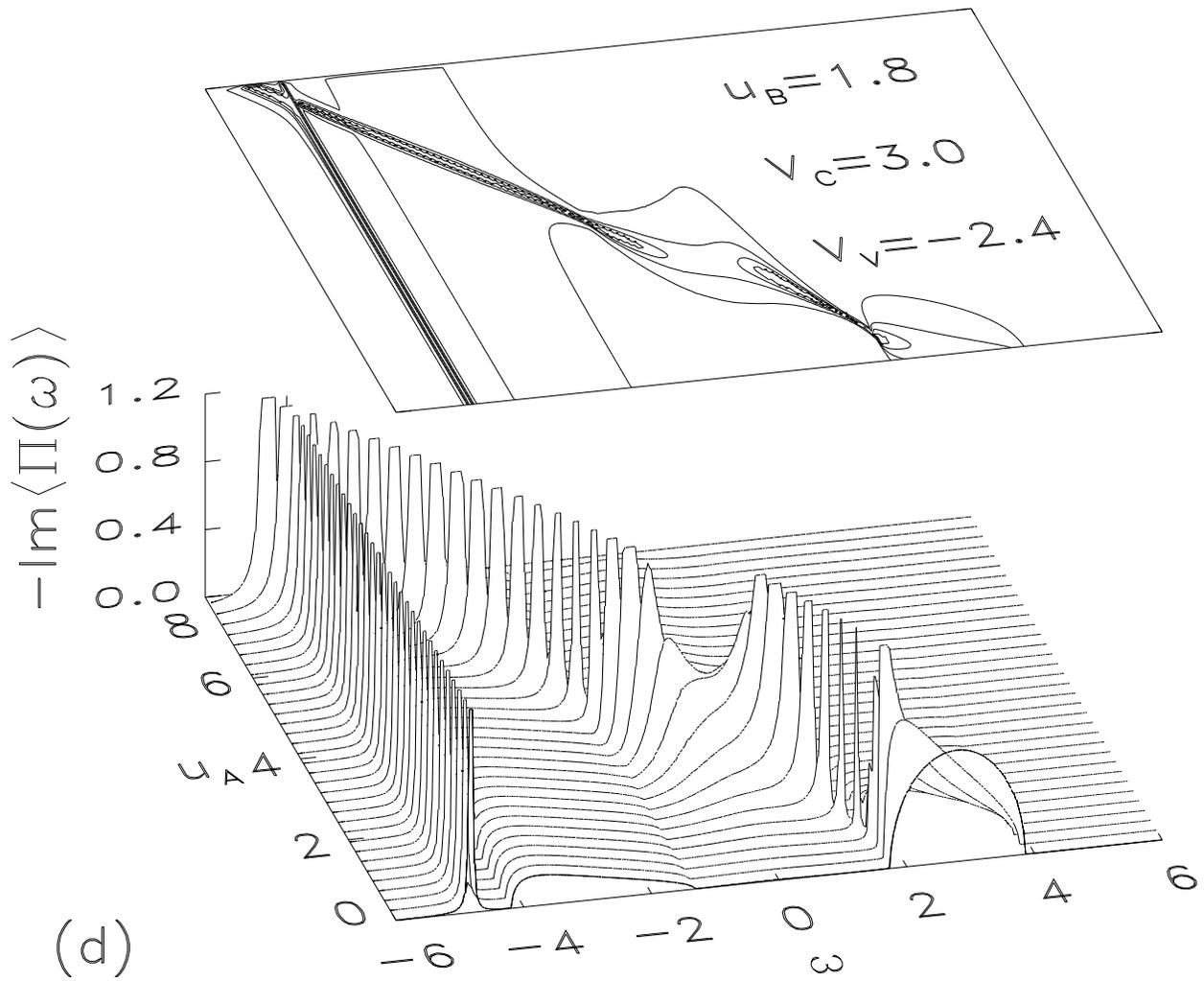
\begin{figure}[tbp]
\caption{Excitonic absorption spectra based on the non-interacting
spectrum in Fig. 4. While in each of the plots (a), (b), (c), (d)
 $u_A$
is varied from 0 to 8 to pull the A-resonance below the onset of the
 lowest
$B$ states, $u_B$ is varied as (a) $u_b=0.0$, (b) $u_B=0.4$, (c)
$u_B=0.8$, (d) $u_B=1.8$. Opposite to the parallel case shown in
 Fig. 3 the
interference of the two interactions amounts to a narrowing of the
$B$-type resonance as the $A$-resonance passes it, cf. Fig. 5 (c).
 Both here and for
the cases shown in Fig. 3 this can be understood to originate from the
mixed $AB$-contribution into which both $u_A$ and $u_B$ couple
weakly. Similar to Fig. 3 the contour lines are taken at level
 heights of
0.1, 0.4, 0.8, 1.2, respectively. Note also that an unweighted contact
potential model with an uniform interaction $u$ of equation (24) would
in this case
erroneously predict a trapping of an excitonic resonance in the gap
region where $Re[\bar \Pi_0] = 0$ at about $\omega=0.8$, as
 $u \rightarrow \infty$.}
\label{Fig5}
\end{figure}

In order to show the effect of the various components of the densities
of states on the formation of excitons in a concise way we have
plotted stacks of three dimensional overlays of the obtained
spectra. In every separate stack of the spectra in Figs. \ref{Fig3}
 (a)--(d)
and Figs. \ref{Fig5} (a)--(d) the strength $u_B$ of the contact
interaction on a
$B$-site of the alloy -- here associated with the lower lying
component
-- is kept constant whereby the
interaction strength $u_A$ on an $A$-site -- associated with the
 higher
lying component -- is varied through an interval
of strengths from zero to a value where it is sufficiently large to
pull the exciton below the lowest component of the contributing
unperturbed joint density of states. The same procedure is performed
over all stacks of
plots in Figs. \ref{Fig3} and \ref{Fig5}
as  $u_B$ is gradually increased through a
similar interval of values as $u_A$.

Although, in an experimental situation it would probably be much
easier
 to vary such parameters
of the system as the concentration and to some extent also the
strengths of the disorder by using different substances for the
production of the solid solutions, a strong variation of the
 contact potential strengths  $u_A$ and $u_B$  is better suited for a
theoretical study of the global features of the excitonic
 absorption predicted by our model. As a result we are able to
 analyze the
excitonic absorption through a whole region of the three dimensional
parameter space spanned by $\omega$, $u_A$ and $u_B$. In addition,
since it is difficult to find a view-angle of the resulting
plots which shows all features of the spectra
simultaneously, we have cut off the resonance peaks at an appropriate
value and overlayed a contour representation of the spectra which
renders more detailed information about the
position and the width of the excitonic resonances as the
interaction strengths are increased.

As can already be seen from equation (\ref{eb4}), the
width of the resonances  will be largely determined through the
behavior of
the double weighted components, i.e. through the magnitude of the
imaginary parts of $\bar \Pi_0^{AA}$  and $\bar \Pi_0^{AB} $
as well as $\bar \Pi_0^{BB}$  and $\bar \Pi_0^{AB} $ at
the solutions of Re$[ \bar  \Pi_0^{AA} ] = -1/ \tilde u_A $
and of  Re$[ \bar  \Pi_0^{BB}  ] = -1/ \tilde u_B $,
respectively. This implies that it is possible to have a very sharp
resonance of the exciton peak associated with the higher lying
component of the joint DOS deep within the region where the DOS is
relatively large, but almost entirely consists of the opposite
component. On
the other hand the resonance can be very broad even in a region where
the cumulative DOS is practically zero, due to the presence of finite
double weighted components which mutually cancel out to a great
 extent, once they are summed.

In Fig. \ref{Fig3} (a) the set of spectra commences with the
completely
unperturbed joint DOS for parallel disorder in Fig. \ref{Fig2}.
The A-exciton
which is being pulled out
 at the lower end of the central bulk part, which consists
almost entirely of $AA-$ and $BB$-components, becomes very sharp
as soon as it leaves the pure A-component at about $\omega =
- 0.7$ corresponding to an $A$-interaction strength $
u_A=0.45$. Beyond this value almost the entire oscillator strength of
the $A$-component is found
in the resonance and thus only the pure $B$-contribution remains in
 the
central part as $u_A$ is increased further. It should be noted that
these resonances show a finite width also when they are exterior of
any of the DOS contributions due to a small artificial imaginary part
of about $10^{-3}$ units which has been
added to the energy in order to ensure the correct analyticity
of the quantities involved at the given numerical accuracy. A
major broadening occurs subsequently as the resonance crosses the
lower $AB$-region which sharpens again once it has passed its lower
 end. The
overlayed contour plot suggests that in the region without almost any
$AA$-states, the increase of the excitonic
binding energy is throughout linear with the increase to the
interaction strength $u_A$. However, as the bound state passes
through the
the lower split off flank, which is constituted of about $50\%$ $AB-$
 and
$25\%$ $AA-$ and $BB$-components, respectively, -- see $10 \times$
 enlarged region
in Fig. \ref{Fig2} (b) -- it gets broadened and the propagation
 path of the bound
state seems to attain a parallel shift (of about $\omega = 0.7$)
 with respect to the initial one, corresponding to a constant
addition to the binding energy beyond
the lowest flank of the spectrum. This additional deepening of binding
is found through all stacks of spectra for parallel and anti-parallel
disorder equally.
It can be observed throughout Fig. \ref{Fig3} (b)--(d) that as the
$A$-resonance passes the lower $AB$-flank
most of the states from this region are being absorbed into the
resonances while the same is true for the the upper $AB$-flank of
the spectrum which feels the effect from afar.
A very similar behavior is observed as $u_B$ is increased over the
series of stacked plots and
the last bits of the upper flank get absorbed in this process. Passing
through the lower $AB$-flank also gives a strong broadening which
 accounts
for the persistence of $B$-states which are not influenced
by the $A$-interaction.

The plots also show that the formation of
excitonic states associated with either of the underlying components
of the alloy is largely independent of correlations between the two
interaction strengths $u_A$ and $u_B$, i.e. the formation of bound
states associated with one of the two material components is seen
to be
as good as unaffected by a variation of
interaction strength associated with the complementary component.
The only slight correlation which comes into play is via the mixed
$AB$-components to which both $u_A$ and $u_B$ couple weakly. It can be
observed that as the resonance associated with the
higher lying component passes through the shifted body of the lower
lying one, both this
resonance and the peak width of the absorption edge narrow in the
 case
of parallel disorder Figs. \ref{Fig3} (b)--(d) and broaden in the
 case of anti-parallel
disorder Figs. \ref{Fig5} (b)--(d).

Figs. \ref{Fig5} (a)--(d) for anti-parallel disorder exhibit many
 similar properties
to the ones for the parallel case,
but they also display quite a few novel features. The $A$-resonance
 is strongly
broadened as it is pulled into the center of the gap, which the
non-interacting absorption for this case from  Fig. \ref{Fig4} (b)
exhibits. This broadening stems from the $AA-$ and $BB$-components
which still prevail in this region and
which are compensated by the $AB$-gain contribution -- see $5 \times$
enlarged region of Fig. \ref{Fig4} (b).
It is very curious to observe however, that even after the resonance
has passed to lower energies, there is a finite hump remaining in
the gap center which is subsequently bleached as the $B$-resonance
 also
shifts downwards. The development of this hump can be followed through
Figs. \ref{Fig5} (a)--(d) if one
 looks through the
trough that the $A$-resonance forms in the central region of the gap
onto the residual spectrum visible at the
back. It mainly consists
of $BB$-states which remain in this region
whereas both the $AA-$ and $AB$-components become largely withdrawn.
Looking at the contour plots of this arrangement suggests that,
in addition to the constant increase in binding of the $B$-resonance
 as $u_B$ increases, which is comparable in magnitude to the one
 observed
for parallel disorder as the central $AB$-region
 is passed, the constant of proportionality for the
deepening seems to have increased corresponding to a faster linear
deepening of the A-exciton binding with increasing $u_A$.

The spectra of Figs. \ref{Fig5} (a)--(d) are also particularly
suitable to visualize that a weighted scattering expansion used in
 the context of an
average decoupling and a contact potential model with two
distinguishable interaction strengths renders a much more accurate
prediction for the formation of the excitonic resonances than the
unweighted one of equation (\ref{ea2_2}). Equation (\ref{ea2_2})
 produces
a resonance whenever Re$[\bar \Pi_0] = - 1/u$ and Im$[\bar \Pi_0]$ is
small or zero. In a split band case as shown in Fig. \ref{Fig4}
the real part is seen to have a zero in the gap where Im$[\bar \Pi_0]
\simeq 0$. This means that an increase of the interaction strength $u$
would lead to a bound state being trapped between the split
bands at this zero of Re$[\bar \Pi_0]$ in the limit $u \rightarrow
\infty$.
This behavior of course is clearly wrong, since one can expect that
the binding energy of any occurring excitonic resonance deepens -- in
this case the binding energy of the resonance  associated with the
 higher lying $A$-band -- as
the interaction strength increases. This failure in giving the right
asymptotic description can be ascribed to a misweighting and
overcounting even of scattering processes which contain two particle
correlations only.
Our approximation overcomes this problem to a great extent and it is
void of the erroneous trapping of the resonance, which as expected
passes on to lower energies
and finally appears below the $B$-band as $u_A$ is increased,
independently of the value of $u_B$.

The variation range of the carrier interaction strengths  --
particularly the one for $u_A$ -- which we
have plotted throughout Figs. \ref{Fig3} and \ref{Fig5} is very large
and is unlikely to be found in semiconductor alloys. However,
 mixtures of molecular crystals
which form extremely narrow bands and at the same time have rather
strong band offsets might be candidates for such behavior.

\section{Comparison to earlier work}

Much earlier work on excitons in strongly disordered binary solid
solutions at finite
concentrations focussed on systems where the exciton is tightly
bound with a binding energy greater that the narrow band width. In
this case the Frenkel exciton has been frequently described by a
``single''
particle Green's function based on the theory of molecular
excitons by Davydov\cite{bia3}. One of
the standard assumptions within this framework, which is based on
 the  Davydov theory, is
that the matrix elements in the optical absorption, $m_{{\bf k},{\bf
k}^{\prime}} = \langle c,{\bf k} \mid \hat \varepsilon \cdot {\bf p}
\mid v, {\bf k}^{\prime} \rangle $ are
strongly localized in momentum space, i.e. $m_{{\bf k},
{\bf k}^{\prime}} =
const \times \delta_{{\bf k},{\bf k}^{\prime}} \delta_{{\bf k},0}$
which means that the contributing transitions are not only required to
be vertical due to the approximate absence of total momentum of the
 exciting
photon but also to only occur at one single point in the bands at
${\bf k}={\bf k}^{\prime}=0$ (Davydov
component), due to the momentum of the relative electron-hole motion
being approximated to zero.

 Early treatments of the absorption of
mixed molecular crystals employed the average amplitude approximation
 (AAA)
first introduced by Broude and Rashba \cite{bia8}, which is able
to roughly predict the position and gravity center strengths of the
dominant transitions only. With the
introduction of the CPA which has some important advantages over the
AAA, such as being able to produce an approximation of the actual
shape of
the joint DOS, it became possible to implement the Davydov theory
within a self consistent framework.

This has been pursued by Onodera and
Toyozawa
 \cite{bia2} and
a little later by Hong and Robinson \cite{bia5} who tried to
model their results particularly on observations made in experiments
\cite{bia4,bia7} on naphtalene and anthracene. The approach of all
 these
authors, despite its self consistency, results in serious
restrictions to the predictions their model can render about aspects
of genuine two particle behavior and it
contains no explicit mention of an electron-hole interaction. It
 seems to give
reasonable predictions on some of the features of Frenkel excitons
 in the
strongly insulating substances as
investigated in references \cite{bia4,bia7}, where the calculated
 positions
of the Davydov components are compared with the ones of the
 experimentally
observed peaks in the absorption spectra.

A significant disadvantage of this single particle exciton
theory is that only one ``joint density of states'' bandwidth and
disorder strength can be considered, rather than two independent
values for both of these parameters, associated with the underlying
conduction and valence bands. A behavior of the joint
density of states as displayed for parallel conduction-valence band
 disorder which can lead to the filling of the
central part of the joint DOS by the states arising from
the $AA-$ and $BB$-transitions even in a split band
limit, as shown in Fig. \ref{Fig2}, could for example never be
obtained in such a simple single particle picture. Additionally,
 the binding
energy of the excitons does not relate to any electron-hole
 interaction, which
in a real system determines their relative position with respect
 to the
continuum absorption edge.

The first work which is based on the footing of a proper two particle
theory known to us is by Abe and Toyozawa \cite{bi11} who
employed Velick{\'y}'s two particle CPA\cite{bi10} to calculate the
 absorption in a non-interacting system with Gaussian disorder.
Kanehisa and Elliott \cite{bi2} thereafter seem to be the first
authors to consider both a two particle CPA for the disorder and
a contact interaction as a model for an electron-hole interaction,
 but they were only able to obtain
an asymptotic solution in the regime of weak disorder, using the
average decoupling of the unweighted
 polarizability in (\ref{ea2_2}). As they pointed out,
their results can be regarded as applying to the amalgamation regime
and a comparison was made with experiments on III-V alloys.

One should note that the approaches of
 Refs. \onlinecite{bi11,bi2,bi12} and the
present work make the assumption that the optical matrix
 elements are constant over the single particle bands involved, even
 though only vertical transitions at ${\bf k} = {\bf k}'$ are allowed,
 which is contrasting the Davydov theory used in the earlier
theoretical models of Refs. \onlinecite{bia2,bia5}, where only the
zero momentum component ${\bf k} = {\bf k}'= 0$ is taken.
In a more realistic situation, the appropriate matrix elements
will exhibit a more general ${\bf k}$-dependence, but it does
not present any conceptual difficulty to modify our present
 results to model such cases as well, once
the actual dispersion laws for the conduction and valence bands
 are known.

\section{Conclusion}

The present work is the most extensive to date which treats the
correlated motion of pairs of interacting particles in a disordered
medium. It draws from the extension of the CPA to two particle
propagation developed by Velick{\'y} \cite{bi10} and its extension to
weighted two particle Green's functions given in our earlier paper
\cite{bi12}.

The model introduced here achieves a better but still approximate
 treatment of
the interference between the effects of impurity scattering and
 direct
two particle (electron-hole) interaction and it provides an
 interpolation scheme
through all strengths of disorder as well as the possibility to
include site dependent variations in the direct electron-hole
interaction within an ATA-like framework. It is therefore most
appropriate for systems where the disorder causes a small but distinct
band splitting, where previous asymptotic models fail to work.

There are many disordered
 systems where such interacting
pairs of particles play an important physical role and we believe the
method can be extended to such situations. Modeling the
excitonic absorption provides a particularly obvious and parmount
 application.

\section{Acknowledgements}

The authors would like to thank Prof. M.A. Kanehisa at Orsay
 Mr. C. Heide at Oxford for reading the manuscript and making
 useful suggestions.

\end{document}